\documentclass[a4paper,fleqn]{cas-dc}

\usepackage[authoryear,longnamesfirst]{natbib}
\usepackage{amsmath} % For math environments
\newcommand{\kms}{\,km\,s$^{-1}$}

\newcommand{\tioi}{$\rm TiO_1$}
\newcommand{\tioii}{$\rm TiO_2$}
\newcommand{\nad}{$\rm NaD$}
\newcommand{\nai}{$\rm NaI8190$}

\newcommand{\cat}{$\rm CaT$}

\newcommand{\hb}{$\rm H_\beta$}
\newcommand{\hg}{$\rm H_\gamma$}
\newcommand{\hd}{$\rm H_\delta$}

\def\tsc#1{\csdef{#1}{\textsc{\lowercase{#1}}\xspace}}
\tsc{WGM}
\tsc{QE}
\begin{document}
\let\WriteBookmarks\relax
\def\floatpagepagefraction{1}
\def\textpagefraction{.001}

\shorttitle{The stellar IMF at $\rm z > 1$ with SHARP}    

\shortauthors{F. La Barbera et al.}  

\title [mode = title]{Probing IMF Variations in High-Redshift Early-Type Galaxies with SHARP}  

\tnotemark[1] 

\tnotetext[1]{This paper is based on a presentation discussing the SHARP science case for the ESO Extremely Large Telescope (E-ELT).} 

\author[1]{F. La\ Barbera}[orcid=0000-0003-1181-6841]

\cormark[1]

\fnmark[1]

% Email id of the first author (placeholder email)
\ead{francesco.labarbera@inaf.it}

% Credit authorship (example based on standard roles for a presentation)
%\credit{  Conceptualization of this study, Methodology, Writing - Original Draft, Data interpretation from related works
%}

\affiliation[1]{organization={INAF-Osservatorio Astronomico di Capodimonte},
           addressline={Via Moiariello 16}, 
           city={Napoli},
           postcode={I-80131}, 
           country={Italy}}
\affiliation[2]{organization={INAF-Osservatorio Astronomico di Trieste},
               addressline={Via Tiepolo 11}, 
               city={Trieste},
               postcode={I-34143}, 
               country={Italy}}
\affiliation[3]{organization={IFPU - Institute for Fundamental Physics of the Universe},
                addressline={Via Beirut 2},
                city={Trieste},
                postcode={34151}, 
                country={Italy}}
\affiliation[4]{organization={INAF-Osservatorio Astronomico di Brera},
               addressline={Via Brera 28}, 
               city={Milano},
               postcode={I-20121}, 
               country={Italy}}
\affiliation[5]{organization={INAF–Istituto di Astrofisica Spaziale e Fisica Cosmica (IASF-Milano)},
               addressline={Via A. Corti 12}, 
               city={Milano},
               postcode={I-20133}, 
               country={Italy}}
\affiliation[6]{organization={INAF-Osservatorio Astrofisico di Arcetri},
               addressline={Largo Enrico Fermi 5}, 
               city={Firenze},
               postcode={I-50125}, 
               country={Italy}}
\affiliation[7]{organization={Instituto de Astrof{\'i}sica de Canarias (IAC)},
               addressline={C/ V{\'i}a L{\'a}ctea, s/n}, 
               city={La Laguna, Tenerife},
               postcode={E-38205}, 
               country={Spain}}

% Co-authors (include key collaborators)
\author[2,3]{G. De Lucia}%[<options>]
\author[4]{F. Ditrani}%[<options>]
\author[2,3]{F. Fontanot}%[<options>]
\author[5]{P. Franzetti}%[<options>]
\author[6]{A. Gallazzi}%[<options>]
\author[5]{A. Gargiulo}%[<options>]
\author[4]{M. Longhetti}%[<options>]
\author[4]{P. Saracco}%[<options>]
\author[1]{C. Tortora}%[<options>]
\author[7]{A. Vazdekis}%[<options>]
\author[6]{S. Zibetti}%[<options>]
%\author{et al-}
% Add more authors as needed from the research group, e.g., Mart{\'i}n-Navarro, Falcón-Barroso

%\affiliation[2]{organization={Instituto de Astrof{\'i}sica de Canarias (IAC) and Departamento de Astrof{\'i}sica, Universidad de La Laguna},
%           city={Tenerife},
%           country={Spain}}
           
%\affiliation[3]{organization={Mullard Space Science Laboratory (MSSL), University College London},
%           city={Dorking, Surrey},
%           country={UK}}

\cortext[1]{Corresponding author}

%\fntext[1]{This research makes use of the SHARP project.}

\begin{abstract}
The stellar initial mass function (IMF), which describes the
distribution of stellar masses at birth, is a fundamental ingredient
in shaping galaxy evolution.  Recent observations indicate that the
IMF varies between galaxies, depending on their mass, morphology, and
stellar content.  In local early-type galaxies (ETGs), spectroscopy,
dynamics, and lensing reveal bottom-heavy IMFs in dense central
regions, with radial gradients toward a Milky Way-like distribution in
the outskirts.  Yet, the chemical enrichment of massive ETGs implies a
dominant role of massive stars during their early formation phases.
These findings can be reconciled if the IMF evolves over cosmic time
-- initially more top-heavy to enable rapid enrichment, and later
dominated by long-lived, low-mass stars.  Directly measuring the IMF
at $z \gtrsim 1$ is therefore essential to test such time-dependent
IMF scenarios, including variations in the dwarf-to-giant and stellar
mass-to-light ratios.  To date, no direct observational confirmation
of these IMF variations -- or of their physical origin -- has been
obtained.  The SHARP spectrograph on the E-ELT, with unprecedented
spatial resolution and sensitivity compared to facilities such as
JWST, and broader spectral coverage than other E-ELT instruments, will
enable spatially resolved spectroscopy of IMF-sensitive features in
high-redshift ETGs up to $z \!  \sim \!  3$, providing unique insights
into the origin of the non-universal IMF in massive galaxies.
\end{abstract}

% Keywords

\maketitle

% Main text
\section{Introduction: The Non-Universal IMF}
\label{sec:Intro}
{ The stellar initial mass function (IMF), i.e.  the distribution of
  stellar masses at birth within a population, is a cornerstone of
  astrophysics. It sets the stellar mass scale of galaxies, regulates
  the strength of stellar feedback, and shapes the chemical enrichment
  and abundance patterns observed throughout cosmic history. Although
  the IMF has traditionally been assumed to be universal, and similar
  to that measured in the solar neighbourhood, several arguments have
  long suggested that it may depend on the physical conditions of star
  formation. For instance, \citet{Larson:1998} argued that early star
  formation may have been biased towards massive stars, i.e.  more
  top-heavy than the present-day IMF, in order to account for the lack
  of metal-free stars and the rapid chemical enrichment of the
  Universe.  Similar conclusions have been reached from
  chemical-evolution models.  \citet{Matteucci:1994} showed that the
  observed increase of $[{\rm Mg}/{\rm Fe}]$ with galaxy mass can be
  reproduced either by a higher star-formation efficiency in more
  massive systems or by a flatter, more top-heavy IMF, while
  \citet{GibsonMatteucci:1997} argued that abundance constraints from
  the intracluster medium favour IMF slopes flatter than Salpeter in
  elliptical galaxies. More recent work has further connected possible
  IMF variations to the thermodynamic and turbulent state of
  star-forming gas, the Jeans or sonic mass scale, metallicity,
  cosmic-ray heating, and star-formation-rate surface density
  \citep[e.g.][]{Vazdekis1997, Weidner2013, Marks2012, Hopkins2013,
    NarayananDave2013, ShardaKrumholz2022, HennebelleGrudic2024}.
  Independent observational and chemical-evolution constraints in
  strongly star-forming galaxies and high-redshift starbursts have
  also been { interpreted as evidence for top-heavy IMFs
    \citep[e.g.][]{Calura2009, Gall2011, Gunawardhana2011, Calura2014,
      Romano2017, Zhang2018, Palla2020, Yan2019, Yan2021}.}  Overall,
  these results indicate that possible IMF variations may occur across
  different environments, both in star-forming and quiescent galaxies
  \citep{AH:18, Smith:2020}, and may be connected to the physical
  conditions under which stellar populations form. }

In the specific case of early-type galaxies (ETGs), \citet{Capp:12,
  Capp:13} revealed a systematic increase in the stellar mass-to-light
ratio ($\rm M/L$) with galaxy mass, based on detailed dynamical models
of stellar kinematics. Their results indicated a transition from a
Kroupa-like IMF at low velocity dispersion ($\sigma \sim 80$~\kms) to
a Salpeter-like IMF at $\sigma \sim 260$~\kms, a finding corroborated
by independent dynamical analyses \citep[e.g.][]{Thomas:11, WCT:12,
  Dutton:12, Tortora:13, Gargiulo2015}.  Additional constraints come
from strong gravitational lensing, which directly probes the total
projected mass on galaxy scales.  { Strong-lensing studies of
  early-type lenses and bulges probing the baryon-dominated inner
  regions, with enclosed lensing masses typically below
  $\sim10^{12}\,M_\odot$, compare lensing-based total masses with
  stellar masses inferred from spatially resolved photometry and
  generally favour a Milky Way-like IMF (e.g. Kroupa/Chabrier) over a
  Salpeter one \citep{FSW:05, FSB:08, ECross:10}.  }  Yet, several
investigations have found a systematic increase in stellar $\rm M/L$
with galaxy mass \citep{Auger:10, Treu:10, Barnabe:11}, though notable
counterexamples exist \citep{SmithLucey2013, SLC:2015, Leier:2016}.
Overall, the consensus emerging from both lensing and dynamical
approaches points to IMF variations in massive ETGs, with departures
from a Kroupa/Chabri{\'e}r form towards either bottom- or top-heavy
distributions -- both leading to high $\rm M/L$ ratios (either because
of low-mass stars or stellar remnants).

Spectroscopic analyses provide a complementary and independent line of
evidence~\citep[e.g.][]{vDC:10, Ferreras2013, LB:13, Spiniello2014}.
By targeting IMF-sensitive absorption features in ETG spectra -- such
as the gravity-dependent Na I doublet at $\lambda\lambda8183,8195$ Å
(NaI\,8190) and the Wing-Ford band (FeH0.99), which are notably
stronger in cool M-dwarfs than in giants \citep{FaberFrench1980,
  SchiavonFeH:97} -- as well as additional indicators like TiO and Ca
features \citep{Cenarro:2003}, spectroscopy directly probes the
stellar population mix. This approach enables constraints not only on
the overall $\rm M/L$ ratio but also on the relative contribution of
low-mass stars shaping the low-mass end of the IMF, although it
remains insensitive to the short-lived, high-mass end.  In the nearby
Universe, several studies have analyzed IMF-sensitive features in
massive ETGs, consistently finding that the central regions of the
most massive systems host a dwarf-enriched (bottom-heavy) IMF
(e.g.~\citealt{CvD12b, LB:13, Spiniello2014, NMN:15a, LB:19,
  Parikh:2018}).

Despite significant observational and theoretical efforts, the
physical origin of IMF variations in the cores of massive ETGs remains
unclear.  { The existence of compact massive galaxies at high redshift,
including compact star-forming and quiescent systems often referred
to as `blue'' and `red nuggets''~\citep[e.g.][]{Daddi2005,
Trujillo2006, vDokkum2008, Barro2013, Dekel2014, Zolotov2015},
suggests that the central regions of present-day massive ETGs formed
under extreme conditions, with gas masses of order $\gtrsim
10^{11},M_{\odot}$ converted into stars within the inner $\sim
1$--$2$ kpc on a short timescale, of order a dynamical time.}
This implies sustained star-formation rates { of several hundred to
$\sim 1000,M_{\odot},{\rm yr}^{-1}$, comparable to those inferred for
compact star-forming galaxies and intense high-redshift starbursts
\citep[e.g.][]{Daddi2005, Barro2013, Dekel2014}, and hence very high gas
densities, pressures, and turbulent velocities.} Such conditions may affect the IMF by
driving highly supersonic turbulence, promoting fragmentation on small
spatial scales~\citep{Hopkins2013}, and/or by shifting the characteristic
fragmentation mass towards lower stellar masses~\citep{Chabrier2014}. These
mechanisms provide a possible route to the bottom-heavy IMF inferred from
IMF-sensitive absorption features in the central regions of nearby
massive ETGs.

However, the physical picture is not unique. Observations of strongly
star-forming galaxies have also been interpreted as evidence for a
top-heavy IMF, i.e. an excess of massive stars relative to a Milky
Way-like IMF \citep{Gunawardhana2011}.  This apparent tension may
indicate that the IMF is both environment- and time-dependent. The
earliest, most intense star-forming phase could have been
characterized by a top-heavy IMF, { increasing the relative number
  of massive-star progenitors and hence the expected production of
  metals and stellar remnants}, while a later phase could have locked
a large fraction of the remaining gas into low-mass stars
\citep{Vazdekis1996,Vazdekis1997,Dave2008}.  Such a transition may
occur if the energy injected into the interstellar medium during the
starburst phase progressively modifies the thermodynamic conditions
for cloud fragmentation.  For instance, a strong cosmic-ray energy
density at high SFR surface density can increase the Jeans mass in
molecular clouds and favour a top-heavy IMF during the earliest phases
\citep{Papadopoulos2011, Fontanot2018b, Fontanot2024}. A time-varying
IMF has also been proposed to alleviate the high abundance of
UV-bright sources at $z>10$ reported by JWST~\citep{Fontanot2026}.
Another possible route is provided by the integrated galactic initial
mass function (IGIMF) framework, in which the galaxy-wide IMF depends
on the star-formation rate because it results from the sum of the IMFs
of individual star-forming clusters~\citep{WeidnerKroupa2005,
  Weidner2013, Yan2017, Jerabkova2018}.  Taken together, these
mechanisms suggest that different stellar-mass ranges may have been
populated under different physical conditions during the formation of
massive ETG cores. { This focus on galaxy cores is motivated by
  local spatially resolved IMF studies, which show that the
  dwarf-enriched component is strongest in the central $\sim1$--$2$
  kpc of the most massive ETGs and becomes progressively more
  Milky-Way-like at larger radii.}  The IMF measured in the core of a
massive ETG may therefore retain the fossil imprint of multiple
star-formation phases occurring under rapidly evolving physical
conditions.

{ In order to constrain this theoretical framework, detailed IMF
  measurements are needed as a function of both environment and cosmic
  time.  } However, no detailed investigation of the IMF -- either in
galaxy cores or as a function of galactocentric distance -- has yet
been conducted beyond the local Universe.  The only notable exception
is the study by \citet{NMN:15d}, who analyzed a single IMF-sensitive
feature (TiO$_2$) in stacked HST grism spectra of quiescent galaxies
at $\rm z \sim 1$, finding an integrated IMF signal { consistent
  with a bottom-heavy IMF, similar to that inferred for the central
  regions of local massive ETGs. This suggests that any transition
  from a top-heavy to a bottom-heavy phase may have occurred early,
  with the resulting low-mass IMF signature evolving only weakly
  thereafter.}

Current JWST programs are beginning to explore the IMF in galaxies up
to $\rm z \sim 1$, yet studies at higher redshift -- and particularly
those probing IMF radial gradients -- will remain beyond reach, even
with JWST’s capabilities, due to its moderate spatial resolution and
sensitivity (see below).  This unexplored regime is where IMF
measurements will offer the most powerful constraints on the origin
and evolution of IMF variations across galaxies.

Next generation near-infrared (NIR) spectrographs, such as SHARP at
the E-ELT, will revolutionize this field, enabling detailed IMF studies at
redshifts beyond one with a level of precision comparable to that 
achieved in the local Universe.  This paper summarizes
the key observational constraints and theoretical insights on IMF
variations in ETGs and outlines the transformative advances that will
be made possible with the SHARP spectrograph on the E-ELT.
With its unprecedented spatial resolution, sensitivity, and { broad
  spectral coverage ($\lambda = 0.95$--$2.45\,\mu{\rm m}$),} SHARP
will allow us to uncover the physical origins of IMF variations in
massive galaxies.  Throughout the paper, we adopt a $\Lambda$CDM
cosmology with $\Omega_m=0.3$ and $H_0=70$\,km\,s$^{-1}$\,Mpc$^{-1}$.

\section{IMF Variations: Current View from Local Studies}

\subsection{The $\text{IMF}-\sigma_0$ Correlation}
\label{sec:imfsigma}
Early attempts to measure the dwarf-to-giant ratio in the IMF go back
over thirty years \citep[e.g.,][]{Cohen:78, FaberFrench1980,
  Carter:86, Hardy:88, Delisle:92}, but were hampered by small
samples, low signal-to-noise, and uncertain stellar population
models. Using the NIR CaII triplet, \citet{Cenarro:2003} suggested a
bottom-heavy IMF in the most massive ETGs.  Significant progress came
with \citet{vDC:10, vDC:11}, who analyzed the NaI8190 and FeH0.99
absorption features, showing that massive ETGs in the Virgo and Coma
clusters have IMFs substantially heavier than Kroupa or Chabri{\'e}r -- a
result later confirmed for 34 SAURON ETGs with full spectral fitting
and variable-abundance stellar population models \citep{CvD12a,
  CvD12b}.

\begin{figure}
  \centerline{ \includegraphics[width=7cm]{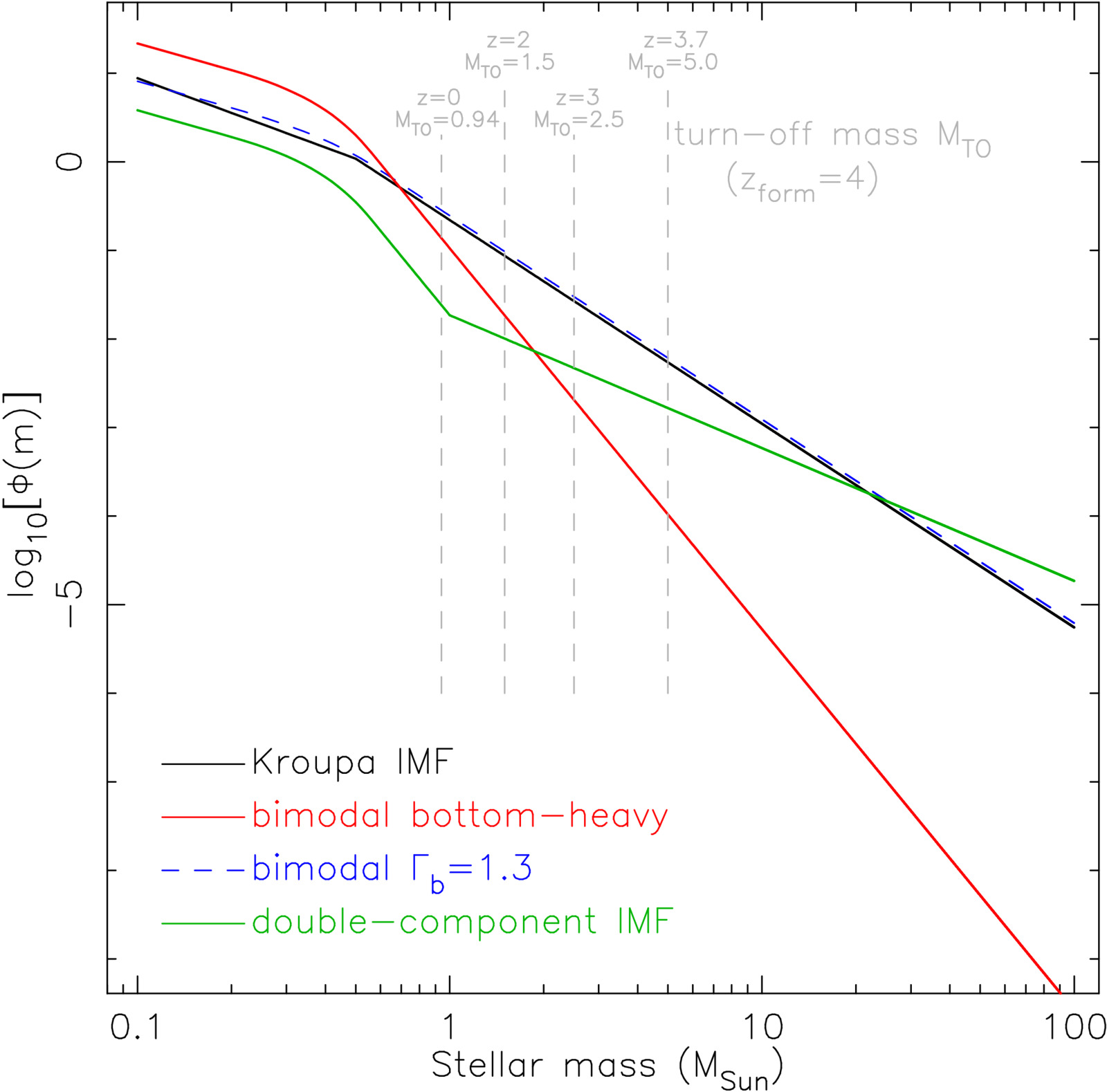} }
  \caption{ The stellar IMF is shown for different parametrizations:
    the Kroupa IMF (black); { a Kroupa-like bimodal IMF, obtained
      for $\Gamma_{\rm b}=1.3$ (blue dashed line; shifted slightly
      upwards for clarity);} a low-mass tapered bottom-heavy bimodal
    IMF (red), with $\Gamma_{\rm b}=3.3$, representative of the
    central regions of massive ETGs at $z\sim0$; and a
    double-component IMF (green), used here only for illustrative
    purposes (see Sec.~\ref{sec:timevar}), created by replacing the
    high-mass portion of the red curve with a flat power-law IMF of
    slope $\Gamma=1.5$ above a stellar mass of $0.97\,{\rm M_\odot}$
    { (see App.~\ref{app:IMF}). All the IMF used in this work are
      defined over the full mass range shown in the plot, namely
      $0.1\leq m/{\rm M_\odot}\leq100$.  For the bimodal IMF,
      $\Gamma_{\rm b}$ denotes the logarithmic high-mass slope, while
      the IMF is smoothly tapered at low stellar masses. Dashed grey
      vertical lines indicate the main-sequence turn-off masses,
      $M_{\rm TO}$, for a stellar population formed at $z_{\rm
        form}=4$ and observed at redshifts $\rm z=0$, 2, 3, and 3.7,
      as labelled.}  }
  \label{fig:IMFshape}
  \end{figure}

\citet[hereafter F13]{Ferreras2013} and \citet[hereafter LB13]{LB:13}
conducted systematic studies of large SDSS ETG samples, analyzing
IMF-sensitive features (Mg, Na, Ca, TiO) using MIUSCAT-MILES stellar
population models \citep{Vazdekis10, Vazdekis:12}. They analyzed 18
stacked spectra spanning $100 \lesssim \sigma \lesssim 300$~\kms, each
with high signal-to-noise, to estimate IMF slopes assuming a low-mass
tapered, or ``bimodal'', IMF.  This parametrization, together with
other IMFs used in the present work, { is shown in
Fig.~\ref{fig:IMFshape}.}  The bimodal IMF is described
by a power-law logarithmic slope, $\Gamma_{\rm b}$, at masses above
$\sim 0.6\,M_{\odot}$, and is tapered towards lower stellar masses.
This allows the dwarf-to-giant ratio to be characterized by a single
parameter, $\Gamma_{\rm b}$, within the adopted stellar-population
models.  We emphasize that, for old stellar populations such as those
in the central regions of ETGs, the spectroscopic features primarily
constrain the integrated contribution of stars with $M \lesssim
1\,M_{\odot}$, rather than the IMF over the full stellar-mass range.
In this parametrization, $\Gamma_{\rm b}=1.3$ provides a Kroupa-like
reference IMF, with a functional form very similar, although not
identical, to that of \citet{Kroupa:2001} over the stellar-mass range
$M \gtrsim 0.1\,M_{\odot}$, i.e.  the lower mass limit adopted in the
stellar-population models used in this work { (see black and blue
  curves in Fig.~\ref{fig:IMFshape}). We note that the Kroupa IMF is
a broken power law, whereas the bimodal IMF is a low-mass tapered
function; the explicit definitions and mass intervals of the IMF
adopted in the present work are summarized in Appendix~\ref{app:IMF}.}

 Figure~\ref{fig:imsigma} shows the IMF slope, $\Gamma_{\rm b}$, as a
 function of $\sigma$ from LB13.  The results indicate a Kroupa-like
 IMF at $\sigma \lesssim 150$~\kms, becoming progressively more
 bottom-heavy at higher velocity dispersion, reaching $\Gamma_b \sim
 2.7$.  { Within the bimodal IMF parametrization,} this bottom-heavy
 slope translates to a mass fraction locked into low-mass stars
 increasing from $\sim 20$ per cent at $\sigma_0 \sim 100$~\kms\ up to
 $\sim 80$ per cent at $\sigma_0 \sim 300$~\kms\ (see LB13 for
 details).

\begin{figure}
  \centering
  \centerline{ \includegraphics[width=8cm]{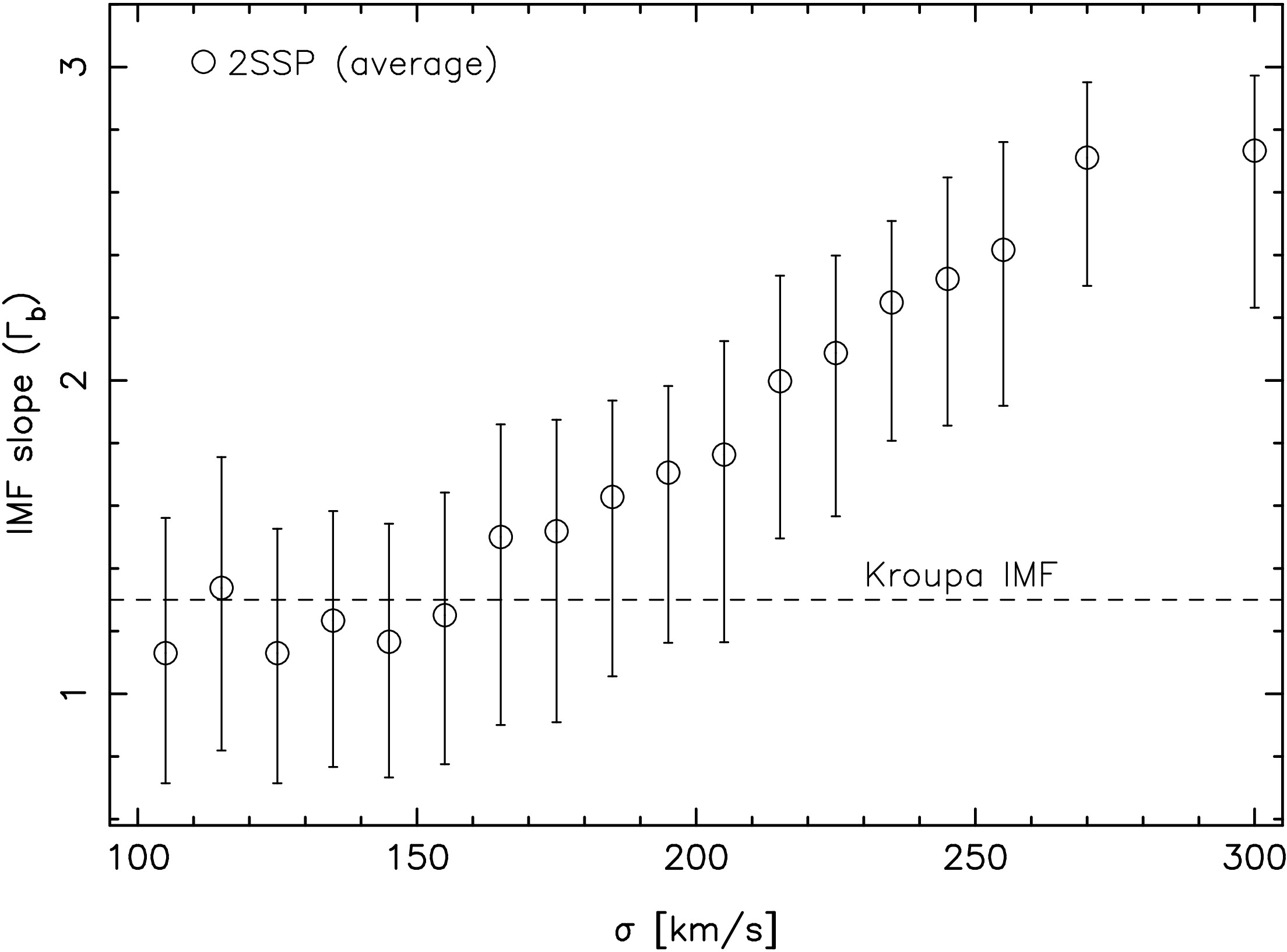} }
  \caption{
    IMF--$\sigma$ relation from~LB13. { The plotted quantity is
    $\Gamma_{\rm b}$, the logarithmic slope of the bimodal IMF at
    $m\gtrsim0.6\,{\rm M_\odot}$; the IMF is smoothly tapered towards
    lower stellar masses. Increasing $\Gamma_{\rm b}$ increases the
    dwarf-to-giant ratio and the contribution of low-mass stars.} Galaxies
    with low velocity dispersion exhibit Kroupa-like IMFs, whereas
    systems with high $\sigma$ show increasingly bottom-heavy IMFs. The
    data points represent the average results obtained from the 2SSP and
    2SSP+XFe spectral-fitting methods in LB13.
  }
  \label{fig:imsigma}
\end{figure}

{ Later work confirmed the IMF--$\sigma$ correlation
  \citep[e.g.][]{Spiniello2014}, showing that this trend does not
  depend critically on the adopted IMF parametrization. However, LB13
  showed that different IMF parametrizations can reproduce similar
  IMF-sensitive spectral features and dwarf-to-giant ratios, while
  predicting significantly different stellar $\rm M_*/L$ values. Thus,
  the conversion from spectroscopic IMF constraints to an IMF
  normalization depends on the adopted IMF functional
  form~\citep{Smith2014}.  At the same time, NIR analyses often report
  weaker or negligible IMF variations \citep{Alton:2017, Alton:2018}.
  One possible reason is that, in old stellar populations, the NIR
  light is dominated by evolved red stars spanning a relatively narrow
  range of initial masses, which may reduce its sensitivity to the
  low-mass end of the IMF.  { Dynamical and lensing studies
    generally support an excess stellar $\rm M/L$ relative to a
    Milky-Way-like, i.e.  Kroupa or Chabri{\'e}r-like, IMF
    normalization \citep{Capp:13,Treu:10}, with exceptions in certain
    lensing systems \citep{Smith2015,Newman2017}.}  However, it is
  important to emphasize that dynamical and lensing studies primarily
  constrain the stellar $\rm M/L$, and that such an excess can in
  principle be produced either by a bottom-heavy IMF, through an
  enhanced fraction of low-mass stars, or by a top-heavy IMF, through
  an enhanced population of stellar remnants (see
  Sec.~\ref{sec:Intro}).}

\subsection{IMF Radial Gradients and Formation History}
\label{sec:IMFgrads}

Early IMF studies in ETGs were limited by integrated spectra, which
average light over large apertures and probe only the global,
luminosity-weighted IMF. { Since integrated estimates of velocity
dispersion, metallicity, age, and elemental abundances are mutually
correlated and also correlate with galaxy mass
\citep[e.g.,][]{Zibetti2020, Gallazzi:21}, a correlation between the
integrated IMF signal and $\sigma$ generally implies correlations with
other stellar-population parameters as well. Integrated measurements
alone therefore make it difficult to identify which physical quantity
is primarily driving the observed IMF variations \citep{LB:15}.}

\begin{figure}
  \centering
  \centerline{ \includegraphics[width=8cm]{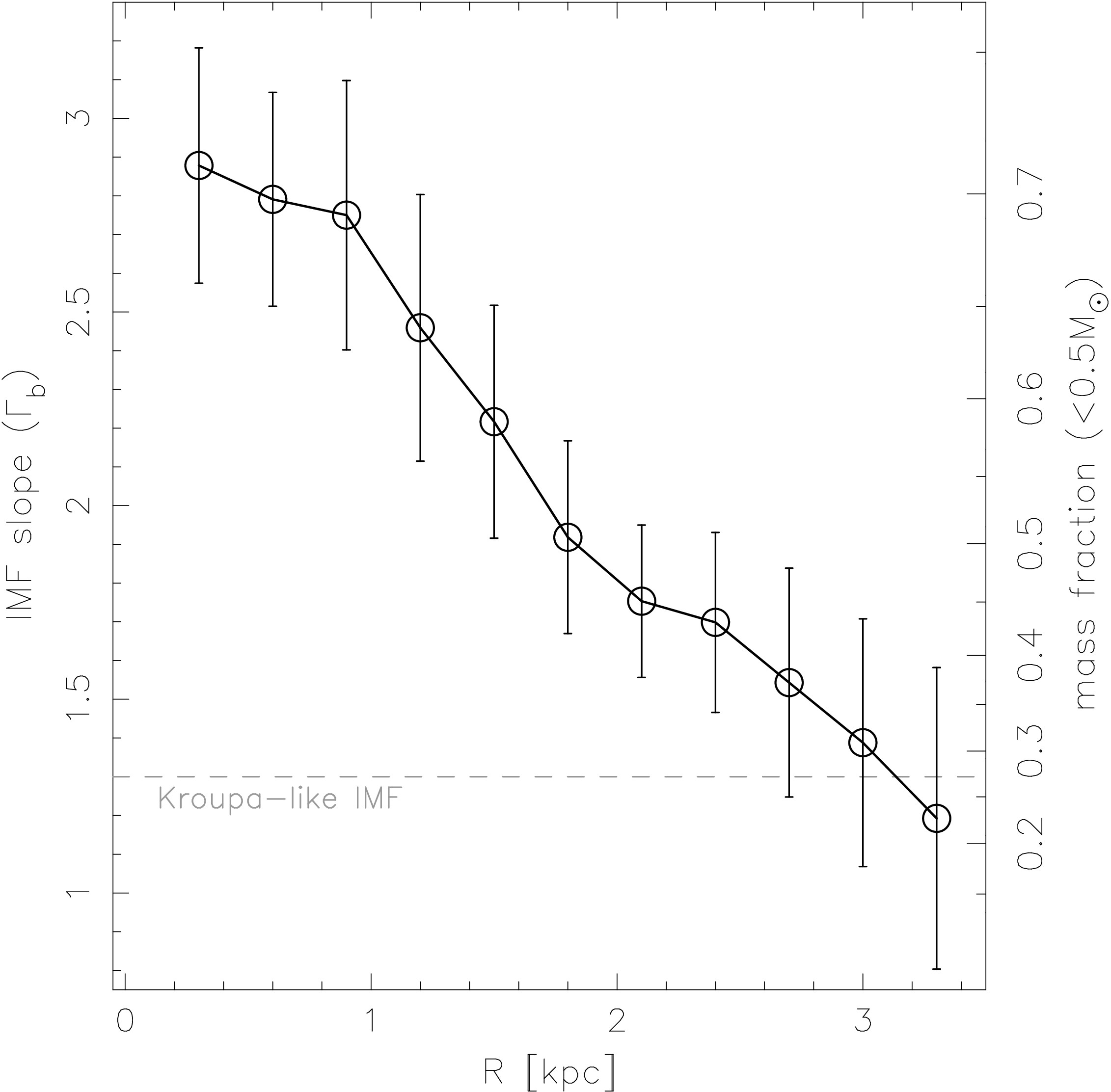} }
  \caption{ Radial IMF gradients for seven massive ETGs from~LB19.
    { The plotted quantity is $\Gamma_{\rm b}$, the logarithmic
      slope of the bimodal IMF at $m\gtrsim0.6\,{\rm M_\odot}$; the
      IMF is smoothly tapered towards lower stellar masses (see
      App.~\ref{app:IMF}).}  The bottom-heavy IMF is concentrated
    within the central $\sim$1–2~kpc and gradually transitions to a
    Milky Way-like distribution in the outer regions. Error bars
    represent the scatter among individual galaxy measurements.  }
  \label{fig:radial}
\end{figure}

Radial IMF gradients were first reported by \citet{NMN:15a}, who found
a gradient in a massive ETG but not in a lower-mass system. Over the
past decade, long-slit and integral-field observations from 10m-class
telescopes have enabled detailed mapping of IMF variations
\citep{NMN:15b, LB:16, LB:17, Ziel:17, vanDokkum:2017, Parikh:2018,
  Sarzi:2018, Dominguez:2019, LB:19, Zhou:2019}. These studies show
that bottom-heavy IMFs are generally confined to the central regions
of the most massive ETGs, while the outskirts resemble the Milky Way.

Figure~\ref{fig:radial} shows the radial profile of the IMF slope,
$\Gamma_{\rm b}$, for seven massive ($\sigma \gtrsim 300$~\kms) ETGs
observed with VLT/X-Shooter \citep[LB19]{LB:19}.  The value of
$\Gamma_{\rm b}$ declines steeply with radius, reaching Kroupa-like
values ($\sim 1.3$) beyond $\sim 2$~kpc. Within this radius, the mass
fraction { locked into stars with $M<0.5\,M_{\odot}$ (see right
  vertical axis in the figure) exceeds $\sim 0.5$. This radius is
  comparable to the typical effective radii of compact high-redshift
  ``red nugget'' galaxies \citep{Damjanov2009}, suggesting that the
  bottom-heavy IMF component observed in local massive ETGs may be
  associated with their dense high-redshift progenitor cores.
  Remarkably, radial IMF profiles are nearly self-similar across
  galaxies when expressed in terms of radius or local surface density
  rather than $\rm R/R_e$, supporting a two-phase formation scenario:
  dense cores form early ($\rm z > 1$) under extreme ISM conditions,
  followed by the accretion, via minor mergers, of lower-density
  stellar envelopes whose stellar populations are consistent with a
  Milky Way-like, or Kroupa-like, IMF \citep{Oser:10,
    NavarroGonzalez2013, vanDokkum2015}.}  Similar trends are seen in
NGC\,1277, a compact relic galaxy~\footnote{A relic galaxy is a system
thought to have formed at high redshift and remained largely
unaffected by subsequent evolutionary processes, such as mergers.}
with a bottom-heavy IMF at all radii \citep[see
  Sec.~\ref{sec:highzgrads}]{NMN:15c}.  Overall, IMF gradient studies
offer a unique probe of how galaxies assemble their stellar mass
through processes such as in-situ star formation and hierarchical
merging.

Despite these advances, the physical drivers of local IMF variations
remain unclear.  TiO$_2$ measurements in CALIFA ETGs suggest
metallicity dominates over local velocity dispersion \citep{NMN:15b}.
%{ This is physically plausible, since metallicity affects gas
%  cooling, dust shielding, and fragmentation, and can therefore modify
%  the characteristic mass scale of star formation.}
However, LB19 found
  that high-metallicity regions in massive ETGs do not always coincide
  with bottom-heavy IMFs.  Analyses of compact stellar systems
  \citep{Villaume:2017, Cheng:2023}, disk-dominated galaxies
  \citep{NMN:19}, and compact ETGs \citep{Maksymowicz-Maciata:2024}
  similarly show that neither metallicity nor velocity dispersion
  alone explains IMF variations.  Theoretically, {
    several physical mechanisms may alter the IMF by changing the
    fragmentation scale of star-forming gas or the relative production
    of low- and high-mass stars. High star-formation rates and gas
    surface densities can modify the galaxy-wide IMF through clustered
    star formation~\citep{Weidner2013}; enhanced cosmic-ray energy
    densities can heat molecular gas and increase the Jeans mass,
    favouring a top-heavy IMF in compact starbursts
    \citep{Papadopoulos2011}; and { metallicity can affect cooling,
    opacity, and fragmentation, modifying the characteristic mass scale of fragmentation, with possible consequences } for the
    low- and high-mass ends of the IMF \citep{Jerabkova2018,
      Chon2024}.}  These results highlight the complexity of the
  problem and the need for multi-dimensional approaches that combine
  kinematics, chemistry, and environment to unravel the origins of IMF
  diversity.

\section{IMF Constraints at redshifts beyond one}

So far, the stellar IMF has not been thoroughly explored beyond the
local Universe.  The redshift domain above one is crucially important,
as direct IMF measurements there would provide the strongest
constraints on the physical origin and cosmic evolution of IMF
variations among galaxies. In the following, we discuss the key
aspects of studying the stellar IMF in high-redshift ETGs, emphasizing
the wavelength range that will be accessible with next-generation
facilities such as the SHARP spectrograph.

\subsection{Redshift range}
\label{sec:wrange}
Constraining the stellar IMF requires the analysis of several spectral
features that are sensitive to different stellar population
parameters. As a representative test case, we consider the following
diagnostic features~\footnote{For each spectral feature, we list the
stellar population parameter to which it is most sensitive. This is a
simplification, as most spectral indices depend on several parameters
simultaneously, but it suffices for illustrative purposes.}:

\begin{itemize}
    \item \textbf{Age:} Balmer lines (\hb, \hg, \hd);
    \item \textbf{Metallicity:} Fe5270;
    \item \textbf{IMF-sensitive features:} Mg4780, \tioi, \tioii, \nad, \nai, and \cat;
    \item \textbf{Abundance ratios:} $C_24668$, $\rm Mgb$.
\end{itemize}

{ The IMF-sensitive features listed above have been widely used in
  local IMF studies because they respond to the relative contribution
  of cool low-mass stars and evolved giants. In particular, \tioi ,
  \tioii , and Mg4780, or bTiO, are sensitive to the effective
  temperatures of the stellar atmospheres contributing to the
  integrated spectrum, and therefore to the dwarf-to-giant ratio
  \citep{Spiniello2014}.  The Na indices are also prominent in the
  atmospheres of dwarf stars; NaI8190 is especially sensitive to very
  low-mass stars, with peak sensitivity at $M \lesssim 0.2\,M_{\odot}$
  \citep{CvD12a}. These features therefore provide useful diagnostics
  of the low-mass end of the IMF.}

\begin{figure}
  \centerline{ \includegraphics[width=8cm]{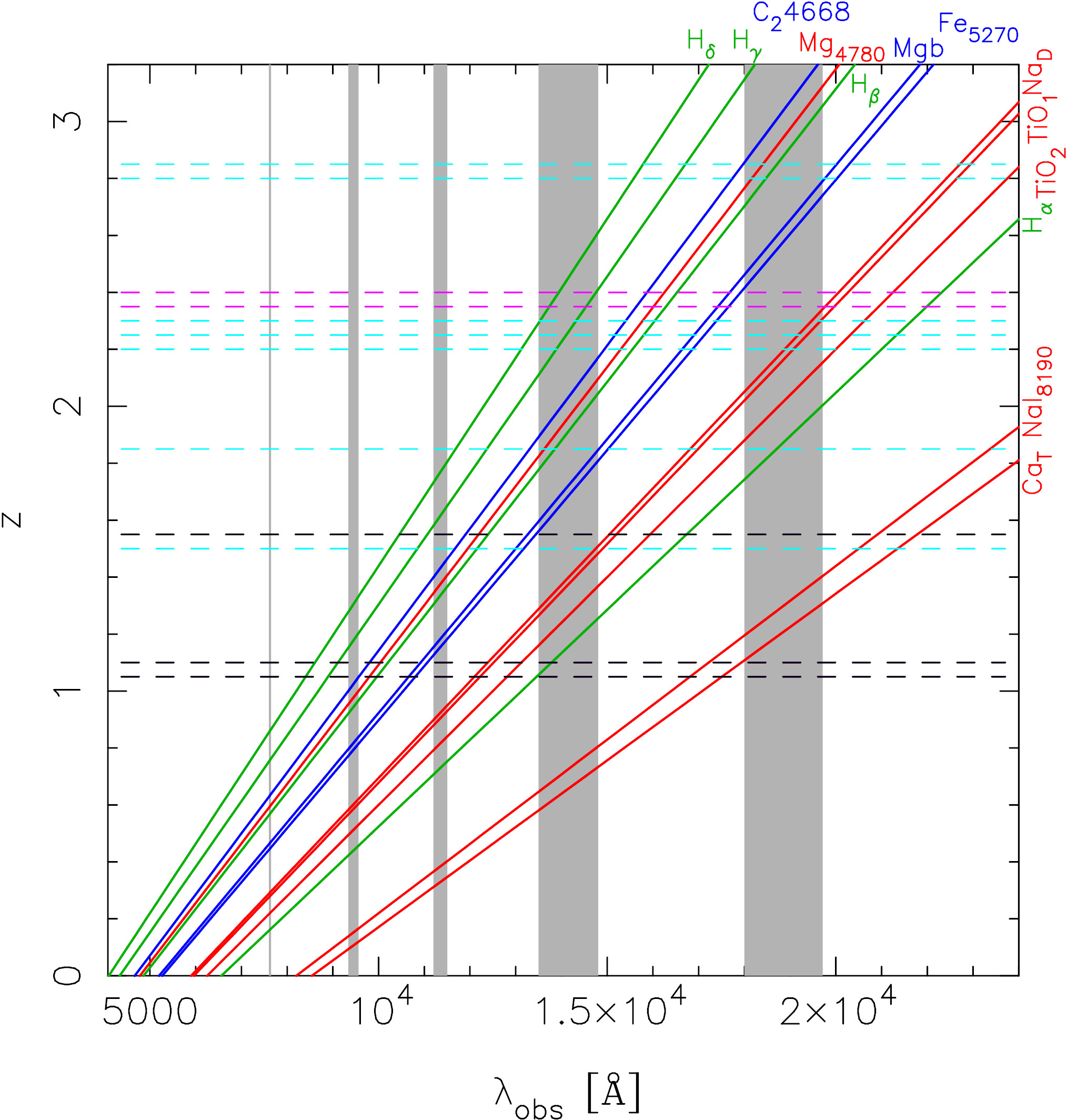} }
  \caption{Observed wavelength as a function of redshift for various
    spectral features: those sensitive to age (green), metallicity and
    abundance ratios (blue), and the IMF (red), as indicated in the
    top-right of the plot (see text for details). Grey vertical bands
    indicate regions affected by strong telluric absorption, where
    spectral features cannot be reliably measured.  Dashed horizontal
    lines mark the redshifts at which age, metallicity, and different
    sets of IMF-sensitive features are accessible: all optical+NIR
    features (black), optical features only (magenta), and TiO features
    (cyan).  }
  \label{fig:wrange}
  \end{figure}

Figure~\ref{fig:wrange} illustrates how these spectral features shift
into the observed frame as a function of redshift, $\rm z$.  Vertical
grey bands mark the positions of the most prominent telluric
absorption regions in the NIR spectral range, while horizontal dashed
lines indicate reference redshifts where the absorption features
remain free from telluric contamination and within the SHARP
wavelength coverage. We consider three cases: (i) all optical+NIR
features (black lines), (ii) optical features only (magenta lines),
and (iii) TiO features only (cyan lines). { Thanks to its spectral
  coverage, $\lambda = 0.95$--$2.45\,\mu{\rm m}$, SHARP will be
  capable of constraining} the IMF up to approximately $\rm z \sim
1.6$ when combining optical and NIR features, $\rm z \sim 2.4$ using
optical features alone, and $\rm z \sim 2.8$ with TiO-based
diagnostics.  It is important to note that SHARP’s capability to
observe in the K band is crucial for comprehensive IMF studies, as it
enables access to a broader set of diagnostic features.  In the
absence of K-band coverage, the corresponding redshift limits would
decrease to approximately $\rm z \sim 1.1$, $1.6$, and $1.8$,
respectively.  Moreover, the K band provides access to powerful
IMF-sensitive features such as FeH0.99 and NaI1.14 -- sensitive to the
very low-mass stars ($\rm < 0.3 M_\odot$) in the IMF -- allowing their
study with SHARP up to redshift $\rm z \sim 1.1$.

\subsection{Constraining the time-varying IMF scenario}
\label{sec:timevar}
Spectroscopic analyses of the IMF in ETGs mainly probe its slope at
stellar masses below the main-sequence turn-off, which in old
populations (i.e.  at $\rm z \sim 0$) corresponds to slightly less
than one solar mass. If the bottom-heavy IMF inferred from these
observations were instead applied to the entire stellar mass range, it
would imply a severe deficit of massive stars -- the primary agents of
metal production and chemical enrichment.  To address this apparent
tension, several authors have proposed that the IMF should vary over
time: an initial, more top-heavy or canonical phase that allows for
efficient enrichment, followed by a subsequent phase dominated by
low-mass, long-lived stars~\citep{Vazdekis1996, Vazdekis1997,
  Weidner2013, Ferreras2015}.  Theoretical studies further support
this picture, suggesting either coupled variations in the IMF at low
and high masses \citep{Fontanot2018b, Fontanot2024, Chabrier2014},
consistent with the proposed time-varying scenario, or modes of
low-mass star formation in which the excess of low-mass stars arises
at later times~\citep{Fabian2024}.  A time-varying IMF scenario
naturally arises within the IGIMF (Integrated Galactic Initial Mass
Function) theory (see \citealt{Jerabkova2018} and references therein),
and has also been proposed to provide better fits of the UV luminosity
function of high-redshift galaxies~(e.g., ~\citealt{Schaerer:2025}).

{ Different parameterizations of the stellar IMF are shown in
  Fig.~\ref{fig:IMFshape} (see Sec.~\ref{sec:imfsigma}):} a Kroupa-like
distribution (black), a bottom-heavy IMF representative of the cores
of massive ETGs at $\rm z \sim 0$ (red), and a double-mode IMF (green)
that is bottom-heavy at low masses and top-heavy at higher masses.
Though the latter is not equivalent to a time-dependent IMF scenario,
it is used here for illustrative purposes as a representation of a
more physically motivated model (see above), where the IMF varies with
time (over a short timescale) from a top- to bottom-heavy phase (or
vice-versa, in the case of a low-mass star-formation mode).  Vertical
dashed lines in the figure indicate the turn-off mass of an old
stellar population, corresponding to a formation redshift of $\rm
z_{form}=4$, typical of massive ETGs, as a function of redshift up to
$\rm z \sim 3.7$.  At $\rm z \sim 0$, the green and red IMFs are
identical { over the mass range of stars that are still alive},
yielding identical spectra -- aside from possible differences in the
abundance pattern -- consistent with observations of the central
regions of massive ETGs.  At higher redshifts, as the stellar
population becomes younger and the turn-off shifts to higher masses,
the two IMFs diverge.

\begin{figure}
  \centerline{ \includegraphics[width=7cm]{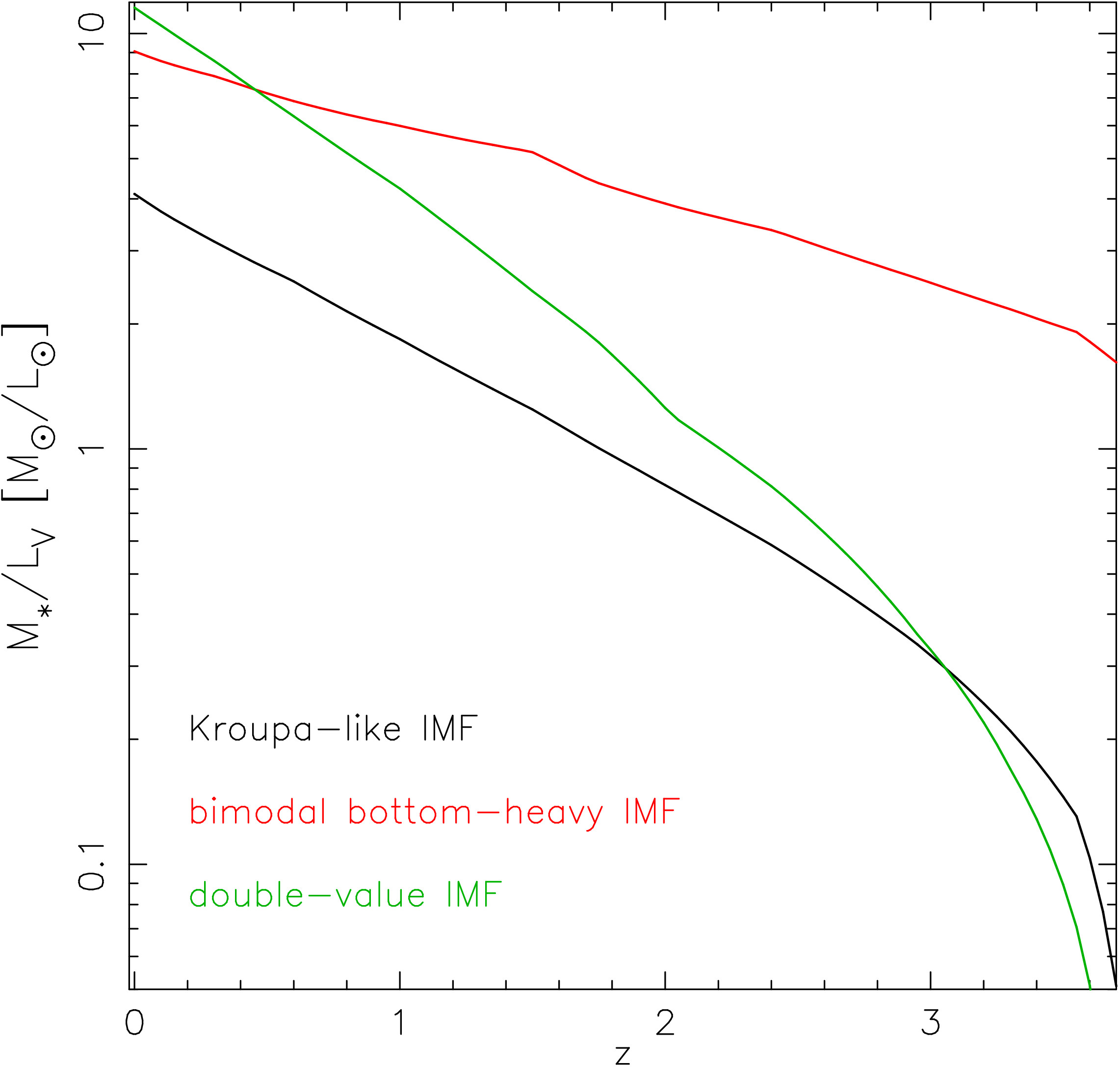} }
  \caption{ Redshift evolution of the r-band stellar mass-to-light
    ratio ($\rm M_{\star}/L_r$) for the different IMF parametrizations
    as in Fig.~\ref{fig:IMFshape}.  In all cases, we assume single
    stellar population (1SSP) models with an old age ($\rm z_{form} =
    4$) and solar metallicity.  }
  \label{fig:ml}
  \end{figure}

Fig.~\ref{fig:ml} shows the time evolution of the stellar
mass-to-light ratio, $\rm M_*/L$, for the same IMFs as in
Fig.~\ref{fig:IMFshape}. { We note that, at redshift $\rm z \sim
  0$, for a fixed low-mass-end slope, the double-mode IMF yields a
  higher $\rm M_*/L$ than the bimodal distribution, because its
  top-heavy high-mass component produces a larger remnant
  contribution. }  Notably, a bottom-heavy IMF produces very high $\rm
M_*/L$ at all redshifts, whereas a double-mode IMF leads to a strong
decline of $\rm M_*/L$ with redshift, decreasing from values typical
of a bottom-heavy IMF at $\rm z \sim 0$ to values consistent with, or
below, the Kroupa IMF at $\rm z \gtrsim 2.5$.  This result is
particularly relevant in light of recent studies of high-redshift
ETGs, which suggest that their dynamical masses, $\rm M_{dyn}$, allow
little room for variations in $\rm M_*/L$ relative to a Kroupa-like
IMF (e.g., \citealt{Forrest2022, Kriek2024}; but
see~\citealt{Belli:2017}).  However, even with JWST, given its spatial
resolution of 100mas/pxl (comparable to the typical effective radius
of ETGs at $\rm z > 1$; see Sec.~\ref{sec:cores}), current
observations cannot yet provide spatially resolved kinematic profiles
of high-redshift ETGs.  Systematic effects, such as rotation, and the
lack of detailed dynamical models may therefore significantly affect
current estimates of $\rm M_{dyn}$~\citep{Slob:2025}.

{ A double-mode IMF can be used to represent a time-varying scenario,
whose physical motivation is discussed in Sec.~\ref{sec:Intro}. This
provides a compelling way to reconcile the} bottom-heavy IMF inferred
at $\rm z \sim 0$ with the relatively low $\rm M_*/L$ observed in
high-redshift ETGs.  This IMF has recently been termed the
``concordance'' IMF~\citep{vanDokkum2024}.  More in general, SHARP
will enable a detailed examination of evolving IMF scenarios, allowing
us to directly test potential time variations in the stellar IMF.

\subsection{IMF-sensitive features vs. redshift}

\begin{figure}
  \centerline{ \includegraphics[width=7cm]{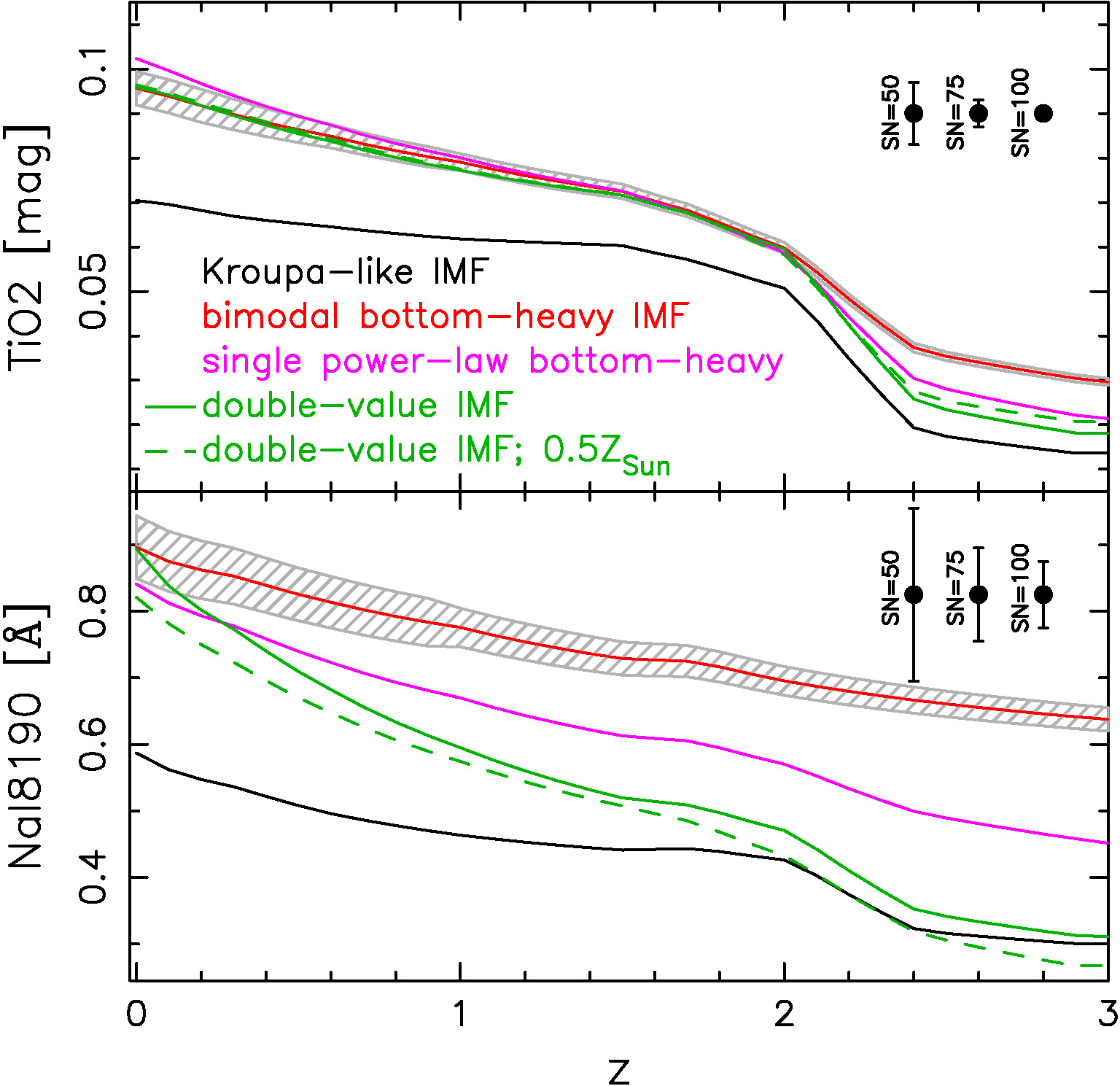} }
  \caption{Redshift evolution of two representative IMF-sensitive
    spectral features, \nai\ (top) and \tioii1 (bottom).  { Solid
      curves with different colors} indicate single stellar population
    (1SSP) models with an old age ($\rm z_{form} = 4$), solar
    metallicity, and the different IMFs shown in
    Fig.~\ref{fig:IMFshape}.  { The dashed curve also shows the
      double-mode IMF model with metallicity 0.2 dex below the
      solar value}.  Hatched grey regions represent the
    expected uncertainties due to errors in elemental abundance
    estimates at each redshift.  Black points with error bars
    illustrate the typical measurement uncertainties for spectra with
    $\rm S/N = 50$, 75, and 100 per \AA .  }
  \label{fig:ews}
  \end{figure}

To explore the possibility of constraining the IMF at high redshift,
Fig.~\ref{fig:ews} shows the redshift evolution of two IMF-sensitive
spectral features, \tioii\ and \nai, for the same IMFs as in
Fig.~\ref{fig:IMFshape}, { plus a single power-law IMF with Salpeter
slope (magenta line). The \tioii\ index responds to variations at the
low-mass end of the IMF because it is sensitive to the effective
temperatures of the stellar atmospheres contributing to the integrated
spectrum, and hence to the dwarf-to-giant ratio. The NaI8190 feature is
especially sensitive to very low-mass stars, with peak sensitivity at
$M \lesssim 0.2\,M_{\odot}$ \citep{CvD12a}.} 
The predictions were generated using E-MILES SSP models with solar
metallicity (solid curves), assuming a single SSP with formation
redshift $\rm z_{form}=4$. { To illustrate the effect of metallicity, the
dashed curve shows the double-mode IMF model with metallicity 0.2~dex
below the solar value adopted in the models. The impact is small for
TiO$_2$, while NaI8190 decreases mildly, by $\lesssim 0.07$~\AA,
relative to the solar-metallicity model.}
Hatched regions in the figure indicate potential uncertainties
($\sim 0.05$~dex) in the IMF-sensitive features arising from errors in
abundance ratios, specifically $\rm [\alpha/Fe]$, $\rm [C/Fe]$, and
$\rm [Na/Fe]$. Black error bars illustrate the expected uncertainties
on the spectral indices for different signal-to-noise ratios, namely
50, 75, and 100 per \AA, respectively. { Index uncertainties were
estimated from 1000 Gaussian-noise realizations at the target S/N,
using the standard deviation of the measured index values.}

The impact of variations in abundance ratios decreases at higher
redshift, making IMF-sensitive features more robust probes of the
stellar mass distribution in younger populations. Among the indices
considered, NaI8190 is particularly effective at distinguishing
between bottom-heavy and double-mode IMF models, even when moderate
uncertainties in abundance ratios are present.  For indices such as
\tioii, the effects of the star formation history (SFH) must be
carefully modeled, as extended or complex SFHs can significantly alter
the predicted feature strengths.

In principle, with high-quality spectra achieving high signal-to-noise
ratios at $z \gtrsim 1$, it should be possible to discriminate between
different IMF scenarios, including bottom-heavy, Kroupa-like, and
double-mode distributions.  Achieving such measurements would allow a
direct investigation of possible time variations in the IMF and
provide crucial observational constraints on the stellar population
properties of massive galaxies in the early Universe.  Moreover,
combining multiple IMF-sensitive features while accounting for
abundance ratio effects and SFH variations will enhance the robustness
of these constraints, enabling a more precise characterization of the
IMF across cosmic time.

\section{Measuring the IMF at High Redshift with SHARP}

SHARP offers an unprecedented opportunity to investigate the stellar
IMF at redshifts beyond unity, during the epoch when the dense cores
of massive ETGs -- believed to host a dwarf-enriched IMF -- were
forming.

\subsection{Probing the Cores of High-Redshift ETGs}
\label{sec:cores}
We consider a representative massive,  quiescent galaxy with a stellar
mass of $\rm  M_\star = 10^{11}\,M_\odot$, observed at  redshifts $z =
1.6$, $2.4$, and $2.8$, corresponding  to regimes where different sets
of IMF-sensitive spectral features are accessible from the ground (see
Sec.~\ref{sec:wrange}).   Each  galaxy  is  modeled  with  a  S\'ersic
surface-brightness profile of index $\rm n = 4$.

Three apparent angular sizes are adopted -- 125, 250, and 500~mas --
corresponding to physical effective radii of approximately 1, 2, and
4~kpc at $z \sim 1$. An effective radius of $\sim 1$~kpc represents a
very compact high-redshift ETG, often referred to as a ``red nugget'',
while 4~kpc corresponds to a larger galaxy, similar to the sizes
observed in massive ETGs at $z \sim 0$.  These values are used as
representative cases at all redshifts.  The slit width is matched to
each apparent size: 70~mas for 125~mas, 140~mas for 250~mas, and
280~mas for 500~mas.  Spectra are assumed to be extracted within an
aperture of $R = R_e/3$, probing the central regions where the IMF is
expected to be most bottom-heavy. H-band magnitudes are set to $\rm H
= 20.8$, 21.2, and 21.6 for galaxies at $z = 1.6$, 2.4, and 2.8,
respectively. Using the SHARP exposure-time calculator (ETC, v.~0.2;
\url{https://sharp.brera.inaf.it/tools/}), we estimate the exposure
time ($T_{\rm exp}$) needed to achieve a restframe signal-to-noise
($\rm S/N$) ratio of 50 per \AA.  The observational setups and
corresponding exposure times are summarized in Table~\ref{tab:texp}.

The required integration times range from a few hours up to $\sim
40$~hrs in the most demanding scenario (the largest apparent-size
galaxy at $z \sim 2.8$). These values are within the limits of typical
observing programs, indicating that SHARP–NEXUS will enable
measurements of the stellar IMF in the central regions of massive ETGs
up to at least $z \sim 3$.  For the VESPER configuration, which has an
overall efficiency $\sim 30\%$ lower than NEXUS, the exposure times
should be scaled by factors of $\sim 1.2$, 1.5, and 1.3 at $z = 1.6$,
2.4, and 2.8, respectively. Although these longer integrations are
more demanding, they remain practical and will allow VESPER to play a
crucial role in mapping radial variations of the stellar IMF, as
discussed in the following section.

\begin{table*}[t]
    \centering
    \caption{Base Exposure Times for SHARP NEXUS Spectroscopy ($\rm
      S/N=50$ per \AA , restframe) to observe massive ETGs at
      redshift z$=1.6$, $2.4$, and $2.8$, respectively.}
    \label{tab:texp}
    \begin{tabular}{ccccc}
        \hline
        Redshift ($z$) & H-band Magnitude & Effective Radius ($R_e$) & Slit Width & Base $T_{\text{exp}}$ \\
        & mag  & mas & mas & hrs \\
        \hline
        1.6 & 20.8 & 125/250/500 & 70/140/280 & 3/8/20 \\
        \hline
        2.4 & 21.2 & 125/250/500 & 70/140/280 & 4/13/29 \\
        \hline
        2.8 & 21.6 & 125/250/500 & 70/140/280 & 6/18/43 \\
        \hline
    \end{tabular}
\end{table*}

\subsection{IMF Radial Gradients at $\rm z > 1$}
\label{sec:highzgrads}
To date, studies  of radial gradients in the stellar  IMF of ETGs have
been  limited to  the nearby  Universe (see  Sec.~\ref{sec:IMFgrads}).
Observations  indicate  that  massive  galaxies  host  a  bottom-heavy
stellar  population  in  their  central regions,  transitioning  to  a
Kroupa-like IMF  at radii larger than  1--2~kpc~\citep{NMN:15a, LB:19}.
While no direct IMF measurements have been performed for high-redshift
progenitors of ETGs (the so-called red nuggets), radial IMF variations
have been  investigated in the  massive galaxy NGC\,1277,  a prototype
relic galaxy  at $\rm z \sim  0$ thought to resemble  ETG progenitors.
In this system, \citet{NMN:15c} found a bottom-heavy IMF at all radii,
consistent with a two-phase formation scenario: the cores of ETGs form
at high  redshift through violent dissipative  processes, producing an
excess of low-mass  stars, while the outer regions  assemble later via
minor  mergers,  exhibiting a  more  standard  IMF.  To  date,  direct
evidence for this two-phase scenario at high redshift remains lacking.

\begin{figure}
  \centerline{ \includegraphics[width=7cm]{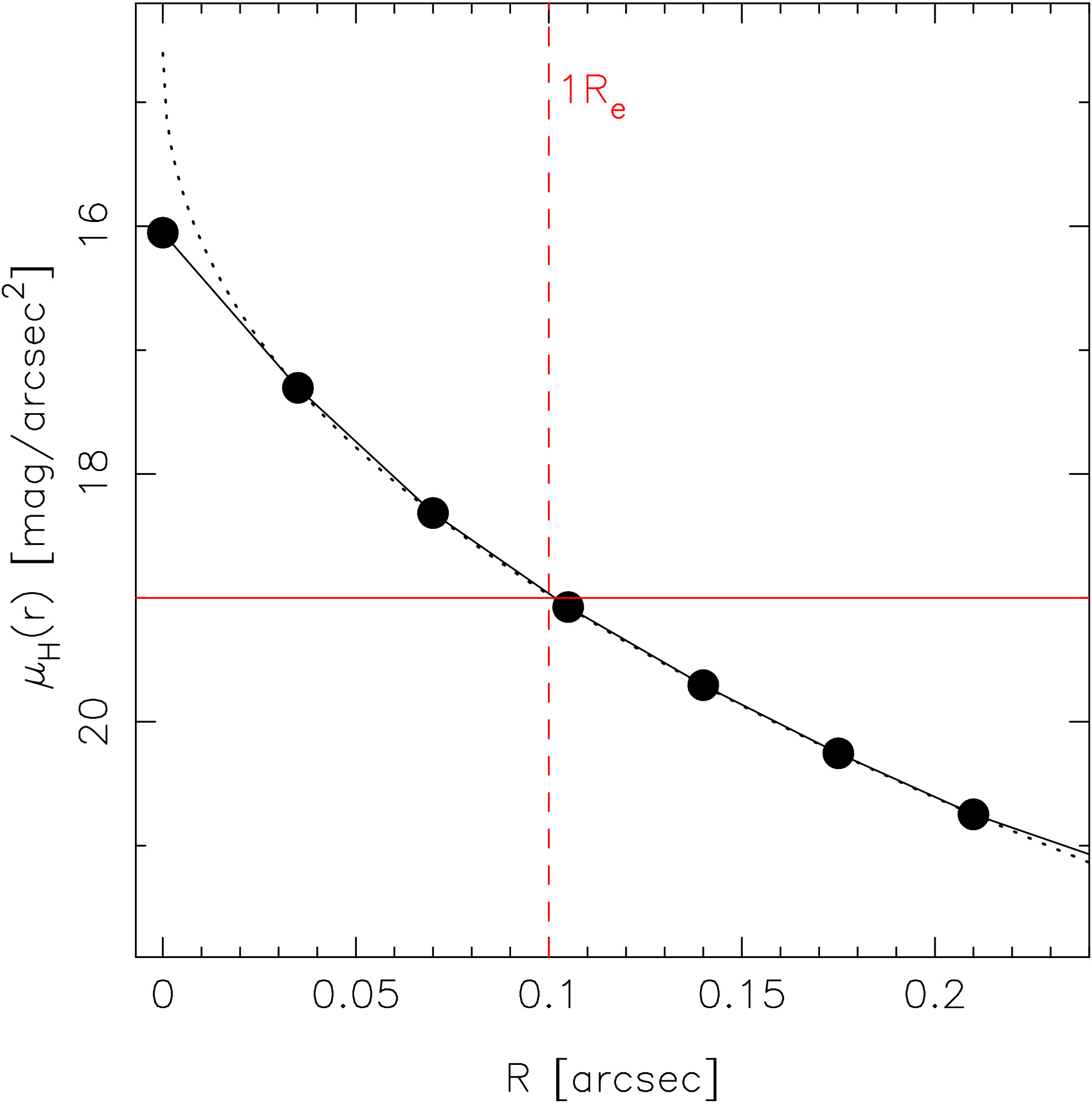} }
  \caption{
Surface brightness profile of a typical compact galaxy at redshift
$\rm z \sim 1.2$, adopting structural parameters from
\citet{Gargiulo2012}. The dotted line shows a S\'ersic profile with
index $\rm n = 2.2$, an effective radius of 100~mas (indicated by the
dashed red vertical line), and a total H-band magnitude of
20.9~mag. The points represent the profile integrated over SHARP–NEXUS
pixels of size 35~mas/px. At $R = R_e$, the galaxy has a surface
brightness of approximately 19~$\rm mag\,arcsec^{-2}$.
  }
  \label{fig:sersic}
  \end{figure}

To illustrate the feasibility of such measurements, we consider a
compact red nugget at $\rm z \sim 1.2$ from the sample of \citet[see
  their Table~1]{Gargiulo2012}, with a S\'ersic index $\rm n = 2.2$,
an effective radius of 100~mas, and a total H-band magnitude of
20.9~mag.  Its surface brightness profile is shown in
Fig.~\ref{fig:sersic}, where points represent the profile integrated
over SHARP–NEXUS pixels of 35~mas/px. At $\rm R = R_e$, the galaxy has
a surface brightness of approximately 19~$\rm mag\,arcsec^{-2}$.  We
consider two observational strategies to measure the IMF at $\rm R =
1\,R_e$: (i) using the NEXUS slit, summing three pixels on both sides
of the slit at $1\,R_e$, and (ii) using a VESPER IFU, summing all
spaxels within an annulus of 0.5--1.0~$R_e$.
{ Based on SHARP ETC calculations, achieving a S/N ratio of $\sim
  50$ per \AA\ at $\rm R=1\,R_e$ requires integration times of 30~hrs
  and 10~hrs for NEXUS and VESPER, respectively. These estimates refer
  to the fainter outer region shown in Fig.~\ref{fig:sersic}; in the
  central regions, which are brighter by $\sim 2$ mag in surface
  brightness, the required exposure times would be shorter by a factor
  of $\sim 6$, i.e. approximately 5~hrs and 1.5--2~hrs for NEXUS and
  VESPER, respectively.}
These estimates demonstrate that SHARP will enable the
study of IMF radial gradients in massive ETGs at $\rm z > 1$, for the
first time, offering critical observational tests of galaxy formation
models. We emphasize that the sensitivity and spatial resolution of
JWST ($\sim 100$~mas) will not allow comparable measurements before
the SHARP era.

\subsection{Leveraging SHARP Multiplexing to Constrain the IMF}
\label{sec:multiplex}

The multiplexing capability of NEXUS–SHARP allows the simultaneous
placement of up to $\sim 30$ slits across a $1.2'\!\times\!1.2'$ field
of view (FOV). Here we explore how this unique feature can be
exploited to constrain the stellar IMF in high-redshift ETGs. To this
end, we estimate the expected number of massive quiescent
galaxies~\footnote{ The progenitors of massive quiescent galaxies in
the local Universe are not necessarily quiescent themselves at $\rm z
> 1$. However, since IMF-sensitive features are difficult to study in
star-forming galaxies due to the dominant light from massive stars, we
focus here, for simplicity, on quiescent progenitors. }, $\rm N_Q$,
within the NEXUS FOV over the redshift range $1 \le z \le 3$, where a
sufficient number of IMF-sensitive absorption features are accessible
(see Sec.~\ref{sec:wrange}).

Given the significant uncertainties in the number densities of massive
galaxies at high redshift, we adopt two independent approaches. Using
number counts from~\citet{Muzzin2013} for field galaxies, we find $\rm
N_Q = 0.4 \pm 0.2$ ($0.9 \pm 0.3$) for stellar masses $\rm M_\star >
10^{11}$ ($10^{10.5}$)~$\rm M_\odot$.  As an independent estimate, we
consider the sample of nine quiescent galaxies with $\rm H < 21.9$ and
redshifts in the range $\rm 1.9 \le z \le 2.3$ presented
by~\citet{Kriek2024}.  These sources were observed across two
MOSFIRE–Keck fields, each covering $6.1'\!\times\!6.1'$, which
corresponds to an expected number of $\rm N_Q = 1 \pm 0.3$ per NEXUS
FOV -- consistent with the estimate derived from~\citet{Muzzin2013}.
We therefore conclude that, for field environments, the multiplexing
advantage of SHARP cannot be fully exploited, as the expected surface
density of massive quiescent galaxies is roughly one per NEXUS field
in the relevant redshift range.

The situation is markedly different in dense environments such as
galaxy clusters. Although such structures are relatively rare at high
redshift, the number of known systems is expected to increase
significantly in the coming years, thanks to ongoing observing
campaigns with JWST and 10~m-class ground-based telescopes.  As a
reference, we consider the rich cluster at $\rm z \sim 1.8$ studied by
\citet{Newman2014}, which includes seven quiescent galaxies with $\rm
H < 21.7$ within a single NEXUS FOV. Given that typical exposure times
required to measure the IMF in such galaxies are $\sim 20$–30~hrs (see
Secs.~\ref{sec:cores} and~\ref{sec:highzgrads}), SHARP’s multiplexing
capability will offer a major advantage over any existing or planned
spectroscopic facilities at the E-ELT or other observatories.  This
will also allow simultaneous, high-quality IMF measurements across
multiple massive galaxies in the same structure, providing
unprecedented constraints on environmental effects in the IMF at high
redshift.

\section*{Summary and Final Remarks}

Local studies of ETGs reveal bottom-heavy IMFs in their dense cores,
transitioning toward Milky Way-like distributions in the outer
regions. Extending these measurements to $z \gtrsim 1$ is essential to
test time-dependent IMF scenarios and to constrain the formation of
the central stellar populations in massive ETGs.

Current JWST programs are beginning to probe the integrated IMF in
galaxies up to $\rm z \sim 1$, yet studies at higher redshift will
remain inaccessible even with JWST’s capabilities. Moreover, given
JWST’s spatial resolution of $\sim$100~mas~px$^{-1}$ -- comparable to
the typical effective radius of ETGs at $\rm z > 1$ (see
Sec.~\ref{sec:cores}) -- it will not be possible to obtain detailed,
spatially resolved kinematic or IMF measurements for these
systems. With its combination of high spatial resolution, sensitivity,
and broad spectral coverage, SHARP will enable spatially resolved
spectroscopy of IMF-sensitive features up to $\rm z \sim 3$.

By contrast, other multi-object spectrographs at the ELT, such as
MOSAIC, will be limited to significantly lower redshifts owing to the
lack of K-band coverage -- reaching only $\rm z < 1.8$ (when using TiO
features only), or $\rm z < 1.1$ (when combining optical and NIR diagnostics)
-- and will not permit spatially resolved studies due to the absence
of adaptive optics correction.  Similarly, while single-unit
integral-field spectrographs such as HARMONI will provide high-quality
spectroscopy, their lack of multiplexing will prevent efficient
studies of the stellar IMF for large galaxy samples (e.g. in cluster
environments) within practical observing times.

To fully exploit SHARP’s capabilities, a dedicated observing program
on the stellar IMF could combine both deep and multiplexed components.
A representative deep campaign would include $\sim$20 high-quality
pointings, probing galaxy cores and radial gradients up to $\rm z \sim
3$, and requiring a total of $\sim$400~hrs (assuming $\sim$20~hrs per
target; see Sec.~\ref{sec:cores}).  This would yield IMF gradient
measurements for a sample comparable in size to current local studies.
In parallel, a multiplexed survey of dense environments (e.g., four
clusters at $\rm z \sim 1$--3) could target $\sim$30 quiescent
galaxies within $\sim$120~hrs (assuming $\sim$30~hrs per pointing and
$\sim$7--8 galaxies per field; see Sec.~\ref{sec:multiplex}),
providing statistical constraints on IMF variations with environment.
Overall, a total investment of $\sim$600~hrs (including $\sim$20\%
overheads) would define a comprehensive legacy program, fully
leveraging SHARP’s sensitivity, spatial resolution, and multiplexing
power to address the origin and evolution of the non-universal
IMF. Such an effort would establish the foundation for a
transformative exploration of IMF variations across cosmic time.

In summary, SHARP will open a new and unique window on the study of
the stellar IMF in massive galaxies, providing direct constraints on
its origin, evolution, and environmental dependence, and offering
unprecedented insights into the formation history of massive galaxies.

\section*{Acknowledgments}
  FLB acknowledges support from INAF 2023 RSN1 Minigrant
  1.05.23.04.01.  FLB acknowledges support from the Fundaci{\'o}n
  Occident and the Instituto de Astrof{\'i}sica de Canarias under the
  Visiting Researcher Programme 2022-2025 agreed between both
  institutions.  FF acknowledges support from INAF 2024 RSN1 Minigrant
  1.05.24.07.01.  ML acknowledges support from INAF 2022 RSN1
  Minigrant 1.05.12.04.01.  PS and the SHARP team acknowledge support
  from the INAF Techno-Grant 2022, SHARP - 1.05.12.02.01, and INAF
  Large-Grant 2024, SHARP - 1.05.24.01.01.  AV acknowledges support from
  grant PID2024-162088NB-100 from the
  Spanish Ministry of Science and Innovation.

%\begin{table}[]
%\caption{Key IMF-sensitive features observed by SHARP (from \cite{LaBarberaSHARP})}
%\label{tbl1}
%\begin{tabular*}{\tblwidth}{@{}LLLL@{}}
%\toprule
%Feature & Rest $\lambda$ ($\text{\AA}$) & Max $z$ in NIR & Constraint \\
%\midrule
%$\text{NaI}8190$ & 8190 & $\sim 1.6$ & Low-mass dwarfs\\
%$\text{TiO}_2$ & $\sim 6500-7500$ & $\sim 2.8$ & Extreme IMF variation\\
%$\text{CaT}$ & $\sim 8500$ & $\sim 1.6$ & Low-mass turnoff\\
%\bottomrule
%\end{tabular*}
%\end{table}

%\section{Summary and Outlook}


\bibliographystyle{cas-model2-names}

\begin{thebibliography}{99}
\bibitem[\protect\citeauthoryear{Alton, Smith, Lucey}{2017}]{Alton:2017} Alton, P.~D., Smith, R.~J., Lucey, J.~R., 2017, MNRAS, 468, 1594
\bibitem[\protect\citeauthoryear{Alton, Smith, Lucey}{2018}]{Alton:2018} Alton, P.~D., Smith, R.~J., Lucey, J.~R., 2018, MNRAS, 478, 4464
\bibitem[\protect\citeauthoryear{Auger et al.}{2010}]{Auger:10} Auger, M.~W., Treu, T., Bolton, A.~S., et al., 2010, ApJ, 724, 511
\bibitem[\protect\citeauthoryear{Barnab{\'e} et al.}{2011}]{Barnabe:11} Barnab{\'e}, M., Czoske, O., Koopmans, L.~V.~E., et al., 2011, MNRAS, 415, 2215
\bibitem[\protect\citeauthoryear{Barro et al.}{2013}]{Barro2013} Barro, G., et al., 2013, ApJ, 765, 104
\bibitem[\protect\citeauthoryear{Belli et al.}{2017}]{Belli:2017}
Belli, S., et al., 2017, ApJ, 834, 30
\bibitem[\protect\citeauthoryear{Bensby et al.}{2017}]{Bensby:17} Bensby, T., et al., 2017, A\&A, 605, A89
\bibitem[\protect\citeauthoryear{Calura \& Menci}{2009}]{Calura2009} Calura F., Menci N., 2009, MNRAS, 400, 1347
\bibitem[\protect\citeauthoryear{Calura et al.}{2014}]{Calura2014} Calura F., Gilli R., Vignali C., Pozzi F., Pipino A., Matteucci F., 2014, MNRAS, 438, 2765
\bibitem[\protect\citeauthoryear{Cappellari et al.}{2012}]{Capp:12} Cappellari, M., et al., 2012, Nature, 484, 485
\bibitem[\protect\citeauthoryear{Cappellari et al.}{2013}]{Capp:13} Cappellari, M., et al., 2013, MNRAS, 432, 1862
\bibitem[\protect\citeauthoryear{Carter, Visvanathan, Pickles}{1986}]{Carter:86} Carter, D., Visvanathan, N., Pickles, A.~J., 1986, ApJ, 311, 637
\bibitem[\protect\citeauthoryear{Cenarro et al.}{2003}]{Cenarro:2003} Cenarro, A.~J., Gorgas, J., Vazdekis, A., et al., 2003, MNRAS, 339, L12
\bibitem[\protect\citeauthoryear{Chabri{\'e}r et al.}{2014}]{Chabrier2014} Chabri{\'e}r, G., Hennebelle, P., Charlot, S., 2014, A\&A, 577, A123
\bibitem[\protect\citeauthoryear{Cheng et al.}{2023}]{Cheng:2023} Cheng, C.~M., et al., 2023, MNRAS, 526, 4004
\bibitem[\protect\citeauthoryear{Chon et al.}{2024}]{Chon2024} Chon, S., Hosokawa, T., Omukai, K., Schneider, R., 2024, MNRAS, 530, 2453
\bibitem[\protect\citeauthoryear{Cohen}{1978}]{Cohen:78} Cohen, J.~G., 1978, ApJ, 221, 788
\bibitem[\protect\citeauthoryear{Conroy \& van Dokkum}{2012a}]{CvD12a} Conroy, C., van Dokkum, P., 2012a, ApJ, 747, 69
\bibitem[\protect\citeauthoryear{Conroy \& van Dokkum}{2012b}]{CvD12b} Conroy, C., van Dokkum, P., 2012b, ApJ, 760, 71
\bibitem[\protect\citeauthoryear{Daddi et al.}{2005}]{Daddi2005} Daddi, E., et al., ApJ, 626, 680
\bibitem[\protect\citeauthoryear{Damjanov et al.}{2009}]{Damjanov2009} Damjanov, I., McCarthy, P.~J., Abraham, R.~G., et al., 2009, ApJ, 695, 101
\bibitem[Dav{\'e}(2008)]{Dave2008} Dav{\'e}, R. 2008, MNRAS, 385, 1470
\bibitem[\protect\citeauthoryear{Dekel \& Burkert}{2014}]{Dekel2014} Dekel, A., Burkert, A., 2014, MNRAS, 438, 1870
\bibitem[\protect\citeauthoryear{Delisle \& Hardy}{1992}]{Delisle:92} Delisle, S., Hardy, E., 1992, AJ, 103, 711
\bibitem[\protect\citeauthoryear{Dom{\'i}nguez et al.}{2019}]{Dominguez:2019} Dom{\'i}nguez S{\'a}nchez, H., Bernardi, M., Brownstein, J.~R., et al., 2019, MNRAS, 489, 5612
\bibitem[\protect\citeauthoryear{Dong et al.}{2018}]{Dong:2018} Dong, H., Olsen, K., Lauer, T., et al., 2018, MNRAS, 478, 5379
\bibitem[\protect\citeauthoryear{Dutton, Mendel \& Simard}{2012}]{Dutton:12} Dutton, A.~A., Mendel, J.~T., Simard, L., 2012, MNRAS, 422, 33
\bibitem[\protect\citeauthoryear{Eftekhari, Vazdekis, La Barbera}{2021}]{elham:2021} Eftekhari, E., Vazdekis, A., La Barbera, F., 2021, MNRAS, 504, 2190
\bibitem[\protect\citeauthoryear{Eftekhari et al.}{2022a}]{elham:2022a} Eftekhari, E., La Barbera, F., Vazdekis, A., et al., 2022a, MNRAS, 512, 378
\bibitem[\protect\citeauthoryear{Eftekhari et al.}{2022b}]{elham:2022b} Eftekhari, E., La Barbera, F., Vazdekis, A., Beasley, M., 2022b, MNRAS, 515, L56
\bibitem[\protect\citeauthoryear{Faber \& French}{1980}]{FaberFrench1980} Faber, S.~M., French, H.~B., 1980, ApJ, 235, 405
\bibitem[\protect\citeauthoryear{Fabian et al.}{2024}]{Fabian2024} Fabian, A. C., Sanders, J. S., Ferland, G. J., McNamara, B. R., Pinto, C., Walker, S. A., 2024, MNRAS, 531, 267
\bibitem[\protect\citeauthoryear{Ferreras, Saha, Williams}{2005}]{FSW:05} Ferreras, I., Saha, P., Williams, L.~L.~R., 2005, ApJ, 623, 5
\bibitem[\protect\citeauthoryear{Ferreras, Saha, Burles}{2008}]{FSB:08} Ferreras, I., Saha, P., Burles, S., 2008, MNRAS, 383, 857
\bibitem[\protect\citeauthoryear{Ferreras et al.}{2010}]{ECross:10} Ferreras, I., Saha, P., Leier, D., et al., 2010, MNRAS, 409, L30
\bibitem[\protect\citeauthoryear{Ferreras et al.}{2013}]{Ferreras2013} Ferreras, I., et al., 2013, MNRAS, 429, L15
\bibitem[\protect\citeauthoryear{Ferreras et al.}{2015}]{Ferreras2015} Ferreras, I., Weidner, C., Vazdekis, A., La Barbera, F., 2015, MNRAS, 448, 82
\bibitem[\protect\citeauthoryear{Fontanot et al.}{2018a}]{Fontanot2018a}
Fontanot, F., De Lucia, G., Xie, L., Hirschmann, M., Bruzual, G., Charlot, S., 2018a, MNRAS, 475, 2467
\bibitem[\protect\citeauthoryear{Fontanot et al.}{2018b}]{Fontanot2018b}
Fontanot, F., La Barbera, F., De Lucia, G., Pasquali, A., Vazdekis, A., 2018b, MNRAS, 479, 5678
\bibitem[\protect\citeauthoryear{Fontanot et al.}{2024}]{Fontanot2024} Fontanot, F., La Barbera, F., De Lucia, G., Pasquali, A., Vazdekis, A., 2024, A\&A, 686, A302
\bibitem[\protect\citeauthoryear{Fontanot et al.}{2026}]{Fontanot2026}  Fontanot, F., De Lucia, G., Xie, L., Zibetti, S., La Barbera, F., Cantarella, S., Hirschmann, M., Charlot, S., Bruzual, G., 2026, A\&A, submitted (arXiv:2603.22405)
\bibitem[\protect\citeauthoryear{Forrest et al.}{2022}]{Forrest2022} Forrest, B., et al., 2022, ApJ, 927, 123
\bibitem[\protect\citeauthoryear{Frogel et al.}{1978}]{Frogel:1978} Frogel, J.~A., Persson, S.~E., Aaronson, M., Matthews, K., 1978, ApJ, 220, 75
\bibitem[\protect\citeauthoryear{Frogel, Persson \& Cohen}{1980}]{Frogel:1980} Frogel, J.~A., Persson, S.~E., Cohen, J.~G., 1980, ApJ, 240, 785
\bibitem[\protect\citeauthoryear{Gall et al.}{2011}]{Gall2011} Gall C., Andersen A. C., Hjorth J., 2011, A\&A, 528, A13
\bibitem[\protect\citeauthoryear{Gallazzi et al.}{2021}]{Gallazzi:21} Gallazzi, A.~R., Pasquali, A., Zibetti, S., La Barbera, F., 2021, MNRAS, 502, 4457
\bibitem[\protect\citeauthoryear{Gargiulo et al.}{2012}]{Gargiulo2012} Gargiulo, A., Saracco, P., Longhetti, M., La Barbera, F., Tamburri, S., 2012, MNRAS, 425, 2698
\bibitem[\protect\citeauthoryear{Gargiulo et al.}{2015}]{Gargiulo2015} Gargiulo, A., Saracco, P., Longhetti, M., Tamburri, S., Lonoce, I., Ciocca, F., 2015, A\&A, 573, 110
\bibitem[Gibson \& Matteucci(1997)]{GibsonMatteucci:1997} Gibson, B. K. \& Matteucci, F. 1997, MNRAS, 291, L8
\bibitem[Gunawardhana et al.(2011)]{Gunawardhana2011} Gunawardhana, M. L. P., Hopkins, A. M., Sharp, R. G., et al. 2011, MNRAS, 415, 1647
\bibitem[\protect\citeauthoryear{Hardy \& Couture}{1988}]{Hardy:88} Hardy, E., Couture, J., 1988, ApJ, 325, 29
\bibitem[Hennebelle \& Grudi{\'c}(2024)]{HennebelleGrudic2024} Hennebelle, P. \& Grudi{\'c}, M. Y. 2024, ARAA, 62, 63
\bibitem[Hopkins(2013)]{Hopkins2013} Hopkins, P. F. 2013, MNRAS, 433, 170
\bibitem[\protect\citeauthoryear{Hopkins}{2018}]{AH:18} Hopkins, A.~M., 2018, PASA, 35, 39
\bibitem[\protect\citeauthoryear{Je\v{r}\'abkov\'a et al.}{2018}]{Jerabkova2018}
Je\v{r}\'abkov\'a, T.,  et al., 2018, A\&A, 620, 39
\bibitem[\protect\citeauthoryear{Kriek}{2024}]{Kriek2024} Kriek, M., 2024, ApJ, 966, 36
\bibitem[Kroupa(2001)]{Kroupa:2001} Kroupa, P. 2001, MNRAS, 322, 231
\bibitem[\protect\citeauthoryear{La Barbera et al.}{2013}]{LB:13} La Barbera, F., Ferreras, I., Vazdekis, A., et al., 2013, MNRAS, 433, 3017
\bibitem[\protect\citeauthoryear{La Barbera, Ferreras \& Vazdekis}{2015}]{LB:15} La Barbera, F., Ferreras, I., Vazdekis, A., 2015, MNRAS, 449, L137
\bibitem[\protect\citeauthoryear{La Barbera et al.}{2016}]{LB:16} La Barbera, F., Vazdekis, A., Ferreras, I., et al., 2016, MNRAS, 457, 1468
\bibitem[\protect\citeauthoryear{La Barbera et al.}{2017}]{LB:17} La Barbera, F., Vazdekis, A., Ferreras, I., et al., 2017, MNRAS, 464, 3597
\bibitem[\protect\citeauthoryear{La Barbera et al.}{2019}]{LB:19} La Barbera, F., Vazdekis, A., Ferreras, I., et al., 2019, MNRAS, 489, 4090
\bibitem[\protect\citeauthoryear{La Barbera et al.}{2024}]{LB:24} La Barbera, F., Vazdekis, A., Pasquali, A., et al., 2024, A\&A, 687, 156
\bibitem[\protect\citeauthoryear{La Barbera et al.}{2025}]{LB:25} La Barbera, F., Vazdekis, A., Pasquali, A., et al., 2025, A\&A, 700, 64
\bibitem[Larson(1998)]{Larson:1998} Larson, R. B. 1998, MNRAS, 301, 569
\bibitem[\protect\citeauthoryear{Lecureur et al.}{2007}]{Lec:2007} Lecureur, A., Hill, V., Zoccali, M., et al., 2007, A\&A, 465, 799
\bibitem[\protect\citeauthoryear{Leier et al.}{2016}]{Leier:2016} Leier, D., Ferreras, I., Saha, P., et al., 2016, MNRAS, 459, 3677
\bibitem[\protect\citeauthoryear{Maksymowicz-Maciata et al.}{2024}]{Maksymowicz-Maciata:2024} Maksymowicz-Maciata, M., Spiniello, C., Mart{\'i}n-Navarro, I., et al., 2024, MNRAS, 531, 2864
\bibitem[\protect\citeauthoryear{Maraston}{2005}]{Maraston:2005} Maraston, C., 2005, MNRAS, 362, 799
\bibitem[\protect\citeauthoryear{Maraston et al.}{2020}]{Maraston2020} Maraston, C., Str{\"o}mb{\"a}ck, G., Thomas, D., et al., 2020, MNRAS, 496, 2962
\bibitem[Marks et al.(2012)]{Marks2012} Marks, M., Kroupa, P., Dabringhausen, J., \& Pawlowski, M. S. 2012, MNRAS, 422, 2246
\bibitem[\protect\citeauthoryear{M{\'a}rmol-Queralt{\'o} et al.}{2009}]{Marmol-Queralto2009} M{\'a}rmol-Queralt{\'o}, E., S{\'a}nchez-Bl{\'a}zquez, P., Vazdekis, A., et al., 2009, ApJ, 706, 212
\bibitem[\protect\citeauthoryear{Mart{\'i}n-Navarro et al.}{2015a}]{NMN:15a} Mart{\'i}n-Navarro, I., La Barbera, F., Vazdekis, A., et al., 2015a, MNRAS, 447, 1033
\bibitem[\protect\citeauthoryear{Mart{\'i}n-Navarro et al.}{2015b}]{NMN:15b} Mart{\'i}n-Navarro, I., La Barbera, F., Vazdekis, A., et al., 2015b, ApJ, 806, L31
\bibitem[\protect\citeauthoryear{Mart{\'i}n-Navarro et al.}{2015c}]{NMN:15c} Mart{\'i}n-Navarro, I., La Barbera, F., Vazdekis, A., et al., 2015c, MNRAS, 451, 1081
\bibitem[\protect\citeauthoryear{Mart{\'i}n-Navarro et al.}{2015d}]{NMN:15d} Mart{\'i}n-Navarro, I., et al., 2015d, ApJ, 798, 4
\bibitem[\protect\citeauthoryear{Mart{\'i}n-Navarro et al.}{2019}]{NMN:19} Mart{\'i}n-Navarro, I., La Barbera, F., Vazdekis, A., et al., 2019, MNRAS, 489, 4090
\bibitem[Matteucci(1994)]{Matteucci:1994} Matteucci, F. 1994, A\&A, 288, 57
\bibitem[\protect\citeauthoryear{Muzzin et al.}{2013}]{Muzzin2013} Muzzin, A., et al., 2013, ApJ, 777, 18
\bibitem[Narayanan \& Dav{\'e}(2013)]{NarayananDave2013} Narayanan, D. \& Dav{\'e}, R. 2013, MNRAS, 436, 2892
\bibitem[\protect\citeauthoryear{Navarro-Gonzalez et al.}{2013}]{NavarroGonzalez2013} Navarro-Gonz{\'a}lez, J., Ricciardelli, E., Quilis, V., Vazdekis, A., 2013, MNRAS, 436, 3507
\bibitem[\protect\citeauthoryear{Newman et al.}{2014}]{Newman2014} Newman, A.~B., Ellis, R.~S., Andreon, S., et al., 2014, ApJ, 788, 51
\bibitem[\protect\citeauthoryear{Newman et al.}{2017}]{Newman2017} Newman, A.~B., Smith, R.~J., Conroy, C., Villaume, A., van Dokkum, P., 2017, ApJ, 845, 157
\bibitem[\protect\citeauthoryear{Oser et al.}{2010}]{Oser:10} Oser, L., Ostriker, J.~P., Naab, T., et al., 2010, ApJ, 725, 2312
\bibitem[Palla et al.(2020)]{Palla2020} Palla, M., Calura, F., Matteucci, F., Fan, X. L., Vincenzo, F., Lacchin, E. 2020, MNRAS, 494, 2355
\bibitem[\protect\citeauthoryear{Parikh et al.}{2018}]{Parikh:2018} Parikh, T., Treu, T., Auger, M.~W., et al., 2018, MNRAS, 477, 3954P
\bibitem[\protect\citeauthoryear{Papadopoulos et al.}{2011}]{Papadopoulos2011} Papadopoulos, P.~P., Thi, W.-F., Miniati, F., Viti, S., 2011, MNRAS, 414, 1705 
\bibitem[\protect\citeauthoryear{Rock et al.}{2017}]{Rock2017} Rock, B., Treu, T., Auger, M.~W., et al., 2017, MNRAS, 468, 3456
\bibitem[Romano et al.(2017)]{Romano2017} Romano, D., Matteucci, F., Zhang, Z.-Y., Papadopoulos, P. P., Ivison, R. J. 2017, MNRAS, 470, 401
\bibitem[\protect\citeauthoryear{Rubin et al.}{2025}]{Rubin:2025} Rubin, K.~H.~R., Prochaska, J.~X., Chen, H.-W., et al., 2025, ApJ, 981, 31
\bibitem[\protect\citeauthoryear{Saracco et al.}{2020}]{Saracco2020} Saracco, P., Longhetti, M., Gargiulo, A., et al., 2020, ApJ, 905, 40
\bibitem[\protect\citeauthoryear{Sarzi et al.}{2018}]{Sarzi:2018} Sarzi, M., Spiniello, C., La Barbera, F., et al., 2018, MNRAS, 478, 4084S
\bibitem[\protect\citeauthoryear{Schaerer et al.}{2025}]{Schaerer:2025} Schaerer, D., Guibert, J., Marques-Chaves, R., Martins, F., 2025, A\&A, 693, 15
\bibitem[Sharda \& Krumholz(2022)]{ShardaKrumholz2022} Sharda, P. \& Krumholz, M. R. 2022, MNRAS, 509, 1959
\bibitem[\protect\citeauthoryear{Schiavon et al.}{1997}]{SchiavonFeH:97} Schiavon, R., Barbuy, B., Rossi, S. C. F., et al., 1997, ApJ, 479, 902
\bibitem[\protect\citeauthoryear{Slob et al.}{2025}]{Slob:2025} Slob, M., et al., 2025, A\&A, 702, 110
\bibitem[\protect\citeauthoryear{Smith \& Lucey}{2013}]{SmithLucey2013} Smith, R.~J., Lucey, J.~R., 2013, MNRASL, 434, 1964
\bibitem[Smith(2014)]{Smith2014} Smith, R. J. 2014, MNRAS, 443, L69
\bibitem[\protect\citeauthoryear{Smith, Lucey \& Conroy}{2015}]{SLC:2015} Smith, R.~J., Lucey, J.~R., Conroy, C., 2015, MNRASL, 449, 3441
\bibitem[\protect\citeauthoryear{Smith et al.}{2015}]{Smith2015} Smith, R.~J., Lucey, J.~R., Conroy, C., et al., 2015, MNRAS, 450, 2006
\bibitem[\protect\citeauthoryear{Smith}{2020}]{Smith:2020} Smith, R.~J., 2020, ARA\&A, 58, 577
\bibitem[\protect\citeauthoryear{Spiniello et al.}{2012}]{Spiniello2012} Spiniello, C., Trager, S., Koopmans, L., et al., 2012, ApJ, 753, L32
\bibitem[\protect\citeauthoryear{Spiniello et al.}{2014}]{Spiniello2014} Spiniello, C., Trager, S., Koopmans, L., et al., 2014, MNRAS, 438, 1483
\bibitem[\protect\citeauthoryear{Spiniello et al.}{2021}]{Spiniello:2021} Spiniello, C., La Barbera, F., Tortora, C., et al., 2021, A\&A, 645, A99
\bibitem[\protect\citeauthoryear{Spiniello et al.}{2024}]{Spiniello:2024} Spiniello, C., Saracco, P., La Barbera, F., et al., 2024, MNRAS, 527, 8793
\bibitem[\protect\citeauthoryear{Spinrad \& Taylor}{1971}]{SpinradTaylor:1971} Spinrad, H., Taylor, B.~J., 1971, ApJ, 164, 200
\bibitem[\protect\citeauthoryear{Thomas et al.}{2011}]{Thomas:11} Thomas, J., et al., 2011, MNRAS, 415, 545  
\bibitem[\protect\citeauthoryear{Tortora et al.}{2013}]{Tortora:13} Tortora, C., Romanowsky, A.~J., Napolitano, N.~R., et al., 2013, ApJ, 765, 8
\bibitem[\protect\citeauthoryear{Treu et al.}{2010}]{Treu:10} Treu, T., Auger, M.~W., Koopmans, L., et al., 2010, ApJ, 709, 1195
\bibitem[\protect\citeauthoryear{Trujillo et al.}{2006}]{Trujillo2006} Trujillo, I., et al., MNRAS, 373, L36
\bibitem[\protect\citeauthoryear{Trujillo et al.}{2009}]{Trujillo:2009} Trujillo, I., Cenarro, A.~J., de Lorenzo-C{\'a}ceres, A., et al., 2009, ApJL, 692, L118
\bibitem[\protect\citeauthoryear{van Dokkum et al.}{2008}]{vDokkum2008} van Dokkum, P.~G., et al., 2008, ApJ, 677, L5
\bibitem[\protect\citeauthoryear{van Dokkum \& Conroy}{2010}]{vDC:10} van Dokkum, P.~G., Conroy, C., 2010, Nature, 468, 940
\bibitem[\protect\citeauthoryear{van Dokkum \& Conroy}{2011}]{vDC:11} van Dokkum, P.~G., Conroy, C., 2011, ApJ, 735, 13
\bibitem[\protect\citeauthoryear{van Dokkum et al.}{2015}]{vanDokkum2015} van Dokkum, P., Conroy, C., Villaume, A., et al., 2015, ApJ, 813, 23
\bibitem[\protect\citeauthoryear{van Dokkum et al.}{2017}]{vanDokkum:2017} van Dokkum, P., Conroy, C., Villaume, A., et al., 2017, ApJ, 841, 68V
\bibitem[\protect\citeauthoryear{van Dokkum \& Conroy}{2024}]{vanDokkum2024} van Dokkum, P., Conroy, C., 2024, ApJ, 973, 32
\bibitem[\protect\citeauthoryear{Vazdekis et al.}{1996}]{Vazdekis1996} Vazdekis, A., Casuso, E., Peletier, R.~F., Beckman, J.~E., 1996, ApJS, 106, 307
\bibitem[\protect\citeauthoryear{Vazdekis et al.}{1997}]{Vazdekis1997} Vazdekis, A., Peletier, R.~F., Beckman, J.~E., Casuso, E., 1997, ApJS, 111, 203
\bibitem[\protect\citeauthoryear{Vazdekis et al.}{2003}]{Vazdekis2003} Vazdekis, A., Cenarro, A.~J., Gorgas, J., Cardiel, N., Peletier, R.~F., 2003, MNRAS, 340, 1317
\bibitem[\protect\citeauthoryear{Vazdekis et al.}{2010}]{Vazdekis10} Vazdekis, A., S{\'a}nchez-Bl{\'a}zquez, P., Falc{\'o}n-Barroso, J., et al., 2010, MNRAS, 404, 1639
\bibitem[\protect\citeauthoryear{Vazdekis et al.}{2012}]{Vazdekis:12} Vazdekis, A., Ricciardelli, E., Cenarro, A.~J., et al., 2012, MNRAS, 424, 157
\bibitem[\protect\citeauthoryear{Villaume et al.}{2017}]{Villaume:2017} Villaume, A., Conroy, C., Johnson, B.~D., et al., 2017, ApJ, 850, 127
\bibitem[\protect\citeauthoryear{Wegner et al.}{2012}]{WCT:12} Wegner, G.~A., Corsini, E.~M., Thomas, J., et al., 2012, MNRAS, 144, 78
\bibitem[Weidner \& Kroupa(2005)]{WeidnerKroupa2005} Weidner, C. \& Kroupa, P. 2005, ApJ, 625, 754
\bibitem[Weidner et al.(2013)]{Weidner2013} Weidner, C., Kroupa, P., Pflamm-Altenburg, J., Vazdekis, A., 2013, MNRAS, 436, 3309
\bibitem[Yan et al.(2017)]{Yan2017} Yan, Z., Jerabkova, T., Kroupa, P., \& Vazdekis, A. 2017, A\&A, 607, A126
\bibitem[Yan et al.(2019)]{Yan2019} Yan, Z., Je\v{r}{\'a}bkov{\'a}, T., Kroupa, P., Vazdekis, A. 2019, A\&A, 629, A93
\bibitem[Yan et al.(2021)]{Yan2021} Yan, Z., Je\v{r}{\'a}bkov{\'a}, T., Kroupa, P. 2021, A\&A, 655, A19
\bibitem[Zhang et al.(2018)]{Zhang2018} Zhang, Z.-Y., Romano, D., Ivison, R. J., Papadopoulos, P. P., Matteucci, F. 2018, Nature, 558, 260
\bibitem[\protect\citeauthoryear{Zibetti et al.}{2020}]{Zibetti2020} Zibetti, S., et al., 2020, MNRAS, 491, 3562
\bibitem[\protect\citeauthoryear{Zieleniewski et al.}{2015}]{Ziel:15} Zieleniewski, S., Houghton, R. C. W., Thatte, N., et al., 2015, MNRAS, 452, 597
\bibitem[\protect\citeauthoryear{Zieleniewski et al.}{2017}]{Ziel:17} Zieleniewski, S., Houghton, R. C. W., Thatte, N., et al., 2017, MNRAS, 465, 192
\bibitem[\protect\citeauthoryear{Zhou et al.}{2019}]{Zhou:2019} Zhou, S., Li, C., Bundy, K., et al., 2019, MNRAS, 490, 2124
\bibitem[\protect\citeauthoryear{Zolotov et al.}{2015}]{Zolotov2015} Zolotov, A., et al., 2015, MNRAS, 450, 2327
\end{thebibliography}

% Loading bibliography database (Placeholder for the bibliography file)

\appendix
\section{IMF parametrizations}
\label{app:IMF}

{
In this appendix we summarize the IMF parametrizations used in the
paper. Throughout, we write the IMF in linear form as
\begin{equation}
\Phi(m) \equiv \frac{dN}{dm}.
\end{equation}
Unless otherwise stated, all IMFs in this work are defined over the
stellar-mass interval
\begin{equation}
0.1 \leq m/{\rm M_\odot} \leq 100 .
\end{equation}
The normalization constants are arbitrary in the equations below and
are fixed, in practice, by the adopted total initial stellar mass.

The \citet{Kroupa:2001} IMF is a broken power law. In the mass range
used in this work, it can be written as
\begin{equation}
\Phi^{\rm Kroupa}(m) = A_{\rm K}
\begin{cases}
  m^{-1.3}       & 0.1 \leq m/{\rm M_\odot} < 0.5, \\
  0.5 \, m^{-2.3}  & 0.5 \leq m/{\rm M_\odot} \leq 100 .
\end{cases}
\label{eq:kroupa_imf}
\end{equation}
where the factor $0.5$ ensures continuity at $m=0.5 \, {\rm M_\odot}$.

The unimodal IMF used in the E-MILES stellar-population models is a
single power law,
\begin{equation}
\Phi^{\rm uni}(m) = A_{\rm uni} \, m^{-(\Gamma+1)} ,
\label{eq:unimodal_imf}
\end{equation}
where $\Gamma$ is the logarithmic slope. A Salpeter-like IMF is obtained
for $\Gamma=1.35$, corresponding to $\Phi(m)\propto m^{-2.35}$.

The bimodal IMF, introduced by
\citet{Vazdekis1996,Vazdekis2003}, is a low-mass tapered IMF. It behaves
as a power law at high stellar masses, while it is smoothly tapered
towards lower masses. Following \citet{Vazdekis2003}, we write it
schematically as
\begin{equation}
\Phi^{\rm bim}(m) = A_{\rm bim}
\begin{cases}
0.4^{-\Gamma_{\rm b}} \, m^{-1} & m/{\rm M_\odot} \leq 0.2, \\
p(m) & 0.2 < m/{\rm M_\odot} < 0.6, \\
m^{-(\Gamma_{\rm b}+1)} & m/{\rm M_\odot} \geq 0.6 .
\end{cases}
\label{eq:bimodal_imf}
\end{equation}
Here $p(m)$ is a spline function defined to ensure a smooth transition
between the low- and high-mass regimes. The bimodal function becomes
progressively shallower towards lower masses and is nearly flat in
logarithmic form below $\sim0.2 \, {\rm M_\odot}$. Thus,
$\Gamma_{\rm b}$ controls the dwarf-to-giant ratio within the adopted
stellar-population models, but it should not be interpreted as a single
power-law slope over the full mass interval. A value
$\Gamma_{\rm b}=1.3$ provides a Kroupa-like reference IMF, although the
bimodal IMF and the Kroupa IMF are not mathematically identical: the
former is smoothly tapered at low masses, whereas the latter is a broken
power law.

Finally, the double-component IMF shown in Fig.~\ref{fig:IMFshape} is
used only for illustrative purposes. It is constructed by retaining the
bottom-heavy bimodal form below a transition mass
$m_{\rm tr}=0.97,{\rm M_\odot}$, and replacing the high-mass part with
a shallower power law of logarithmic slope $\Gamma=1.5$ above
$m_{\rm tr}$. In linear form, it can be written as
\begin{equation}
\Phi^{\rm dbl}(m) =
\begin{cases}
\Phi^{\rm bim}(m) & 0.1 \leq m/{\rm M_\odot} < m_{\rm tr}, \\
C \, m^{-2.5} & m_{\rm tr} \leq m/{\rm M_\odot} \leq 100 ,
\end{cases}
\label{eq:double_imf}
\end{equation}
where the constant $C$ is chosen so that the IMF is continuous at
$m_{\rm tr}=0.97,{\rm M_\odot}$. This parametrization is not used as a
fitted IMF model, but only to illustrate a possible time-dependent
scenario in which the low- and high-mass parts of the IMF may have been
set during different phases of star formation.
}
\end{document}